\documentclass{aa}  

\usepackage{graphicx}
\usepackage{txfonts}
\usepackage[toc,page]{appendix}

\usepackage{cancel} 

\begin{document}
\title{Are Over-massive Haloes of Ultra Diffuse Galaxies Consistent with Extended MOND? }

   \subtitle{}

   \author{Alistair O. Hodson
          \and
          Hongsheng Zhao
          }

   \institute{School of Physics and Astronomy, University of St Andrews,
             Scotland\\
              \email{aoh2@st-andrews.ac.uk}
              }

   \date{Received ; accepted }

 
  \abstract
   {}
   {A sample of Coma cluster ultra-diffuse galaxies (UDGs) are modelled in the context of Extended Modified Newtonian Dynamics (EMOND) with the aim to explain the { large} dark matter-like effect observed in these cluster galaxies. }
   {We first build a model of the Coma cluster in the context of EMOND using gas and galaxy mass profiles from the literature. Then assuming the dynamical mass of the UDGs satisfies the fundamental manifold of other ellipticals, and that the UDG stellar mass-to-light matches their colour, we can { verify} the EMOND formulation by comparing two predictions of the baryonic mass of UDGs.}
   {{ We find that EMOND can explain the UDG mass, within the expected modelling errors, if they lie on the fundamental manifold of ellipsoids, however, given that measurements show one UDG lying off the fundamental manifold, observations of more UDGs are needed to confirm this assumption.}}
   {}


   \maketitle
   
\section{Introduction}

Gravitational potential wells of galaxy clusters have been powerful laboratories to test the current model of dark matter ($\Lambda$CDM) and its alternatives.
While acknowledging many shortfalls of the $\Lambda$CDM model in galaxies \citep[e.g.][]{cuspcore1, cuspcore2, missing1, missing2, TBTF1, TBTF2, plane1,Pawlowski}, in the cluster arena few models can compete with the $\Lambda$CDM model, especially those alternatives which a priori assume that particle dark matter does not exist, but rather what we are seeing is a breakdown of Newtonian dynamics.

Observations of rotation curves in galaxies showed that dark matter effects were only required in low acceleration environments $\lessapprox  1.2\times10^{-10}$ ms$^{-2}$. This eventually led to the construction of the empirical gravitational paradigm known as Modified Newtonian Dynamics (MOND) \citep{milgrom19832,milgrom19833,milgrom19831,bekenstein1984}. The main function of MOND is to modify gravity in these low acceleration environments such that the gravitational acceleration falls proportional to $1/r$ in contrast to the Newtonian $1/r^{2}$. Newtonian dynamics are still preserved in the high acceleration environments. In order to achieve this, an acceleration scale was introduced to define what is meant by high and low acceleration environments, $a_{0} \approx 1.2\times10^{-10}$ ms$^{-2}$, such that Newtonian behaviour is recovered when $a>>a_{0}$ and the $1/r$ gravity law (Deep-MOND regime) occurs when $a<<a_{0}$, where $a$ is the total gravitational acceleration.

The MOND paradigm has had success on the galaxy scale, see \cite{famaeyreview} for an extensive review. One of the main problems in MOND is its inability to explain galaxy clusters, see for example \cite[][]{sanders1999,sanders2003}. Galaxy clusters tend to have an internal acceleration of the order $a_{0}$ and thus the MOND effect is weak. However, galaxy clusters show a large mass discrepancy from Newtonian predictions, much more than MOND is able to account for. This means that either { 1) There exists a $\Lambda$CDM dark matter halo, 2) There is missing  matter which we are yet to detect in the form of non-luminous baryonic matter or some form of neutrinos or  3) MOND is not a complete gravity theory and needs to be generalised.} Work on { point 2} has achieved mixed results. \cite{angus20081} have shown that the 2 eV neutrino was insufficient to explain the galaxy cluster problem as their inclusion could not explain mass discrepancy in the centre of the clusters. The neutrino idea was then reinvestigated in \cite{angus2009} where 11 eV sterile neutrinos were tested. This work enjoyed more success in explaining the galaxy cluster problem in MOND and also had success in explaining the CMB anisotropies.  However, cosmological simulations conducted by \cite{angusneutrino1} and \cite{angusneutrino2} showed that using neutrinos as hot dark matter in the MOND paradigm produces too many high mass galaxy clusters. 

A recent addition to the galaxy cluster problem in MOND is the discovery of ultra-diffuse galaxies (UDGs) \citep[][]{UDGComa,UDGA168,UDGVirgo,UDGDF}.  These galaxies have very little gas and are composed almost entirely of dark matter. Recent studies of a UDG in the Virgo cluster \citep{UDGVirgoObject} and the Coma cluster \citep{UDGComaObject} have shown that two UDGs, VC1287 and DF44 show a very high dark-to-stellar mass ratio.

In this work, we are interested in the nature of UDGs in the context of MOND. In a MONDian paradigm, it is possible to create a large dark matter-like effect if the gravitational acceleration across the system is very low. The MOND paradigm has an interesting feature called the external field effect (EFE) (see for example \cite{EFE1}, \cite{EFE2} and \cite{decliningcurves} for the interested reader). The EFE states that even a constant acceleration from an external source can affect the internal dynamics of a system. For example, a stellar cluster located close to the Milky Way disk should behave differently  if it is moved further away from the disk as the gravitational acceleration across the cluster from the Milky Way would be less. In the context of the UDGs, if they were isolated objects, MOND would predict a large dark matter like effect, but as they are within the strong gravitational field of the galaxy cluster, MOND predicts they should behave closer to Newtonian.

Taking this into consideration, if MOND is to be generalised to try and explain the missing mass in galaxy clusters, it must also explain the nature of these UDGs. One modification to MOND which has been proposed is that of Extended MOND (EMOND) \citep{EMOND}. This extension of MOND changes the acceleration scale $a_{0}$ from being constant to being a function of gravitational potential, $A_{0}(\Phi)$, such that the effective acceleration scale in galaxy clusters is much larger than $a_{0}$. This allows deviations from Newtonian dynamics, and hence the inducing of dark matter-like effects, to occur at higher accelerations. We explored EMOND in \cite{HodsonEMOND} with a sample of 12 galaxy clusters. \cite{HodsonEMOND} showed that EMOND has some success with the basic formulation, but no attempt was made to explore the boundary conditions of the gravitational potential to try to get better fits. Also, the exact form of the baryonic mass profile is relevant when determining the EMOND prediction and thus different mass models should be tested in future. As a consequence of this, the paradigm requires more rigorous testing.

Recent work on the UDGs has allowed dynamical mass estimates to be made, that is the total mass of the UDGs including any dark component, for a sample of galaxies from the Coma and Virgo clusters using scaling relations \citep{UDGFM}. This method takes advantage of the fundamental manifold (FM) \citep[][]{FM1,FM2,FM3} to calculate velocity dispersions of the UDGs from their effective radius and surface brightness. The FM is an extension of the fundamental plane \citep[][]{FP1,FP2}. From the velocity dispersions, it is then possible to estimate a dynamical mass for the objects. It is also possible to estimate the stellar mass of the UDGs from their g-i colour. This technique was performed in \cite{UDGComaObject} for DF44 by using the colour-$M_{\star}/L$ correlation from \cite{GAMAStellar}. Therefore it is possible to get both dynamical and stellar mass estimates for a sample of UDGs.

By modelling the Coma cluster in the context of EMOND, we can find the value of $A_{0}(\Phi)$ in the cluster as a function of radius. Assuming this value is constant across any UDG, we can estimate the stellar mass of the UDGs from dynamical mass estimates using the EMOND recipe and compare the result to stellar mass predicted from the colour. By doing this, we can determine whether the EMOND recipe can predict both the mass profile of the Coma cluster and the dynamical to stellar mass fraction of the UDGs simultaneously.

This paper is organised as follows. Section \ref{EMOND} discusses the MOND and EMOND paradigm. Section \ref{Coma} discusses the Coma cluster model which we adopt. Section \ref{UDG} discusses the UDG dynamical and stellar mass estimates from the literature. The UDG modelling in the context of MOND and EMOND is discussed in Section \ref{MOND}.  We show our results in Section \ref{Results}. In Section \ref{refine} we show how the constraints on the EMOND formalism from the UDG modelling affect the results of \cite{HodsonEMOND}. In Section \ref{contention}, we discuss possible contention with observational data. We then conclude in Section \ref{Conclusion}.

\section{Extended MOND}\label{EMOND}

We begin our discussion of EMOND by reviewing the standard MOND equations. In gravitational dynamics, the gravitational acceleration and matter density are linked via a Poisson equation. The MOND Poisson Equation is \citep{bekenstein1984},

\begin{equation}
4\pi G \rho = \nabla \cdot \left[ \mu\left(\frac{|\nabla \Phi|}{a_{0}}  \right) \nabla \Phi \right]
\end{equation}

\noindent where $\rho$ is the matter density and $\Phi$ is the total gravitational potential. The function, $\mu(x)$ is called the interpolation function which models the transition between the Newtonian regime and the Deep-MOND regime. $\mu(x)$ must have limits such that when $x<<1$, $\mu(x) = x$ and when $x>>1$, $\mu(x)= 1$. The form for the interpolation function which we will use in this work is a modified simple interpolation (see \citep{simplemu,simple2} for simple interpolation function)

\begin{equation}
\mu(x)= \rm max \left[\frac{x}{1+x}, \frac{\epsilon}{1 + \epsilon} \right],
\end{equation}

\noindent where $\epsilon$ is a small number. The EMOND version of the MOND Poisson Equation is \citep{EMOND}\footnote{The additional $T_{2}$ term arises from the non-relativistic EMOND Lagrangian. Merely making the change $a_{0} \rightarrow A_{0}(\Phi)$ in the Poisson equation will not satisfy the Euler-Lagrange equation.}

\begin{equation}\label{EMONDPoiss}
4 \pi G \rho = \nabla \cdot \left[\mu\left( \frac{|\nabla \Phi|}{A_{0}(\Phi)}  \right) \nabla \Phi  \right] - T_{2},
\end{equation}
where
\begin{equation}
T_{2} = \frac{1}{8\pi G}\left| \frac{d (A_{0}(\Phi))^{2}}{d\Phi} \right|\left[ y F'(y) - F(y) \right].
\end{equation}
\noindent Also, $dF(y)/dy = \mu(\sqrt{y})$ and $y = |\nabla \Phi|^{2}/A_{0}(\Phi)^{2}$. It was shown explicitly in \cite{HodsonEMOND} that the $T_{2}$ term is negligible in clusters and thus the approximate spherical version of the EMOND Poisson Equation reduces to,

\begin{equation}\label{EMONDPoiss2}
\nabla \Phi_{N} \approx \mu\left( \frac{|\nabla \Phi|}{A_{0}(\Phi)}  \right) \nabla \Phi
\end{equation}
\noindent where $\nabla \Phi_{N}$ is the Newtonian acceleration. The functional form of $A_{0}(\Phi)$ we use here is\footnote{In \cite{HodsonEMOND}, the function for $A_{0}(\Phi)$ was written as $A_{0}(\Phi) = a_{0} + (A_{0~max}-a_{0})\left[ \frac{1}{2}\tanh \left[ \log \left(\frac{\Phi}{\Phi_{0}}  \right)^{q} \right] +\frac{1}{2}\right]$.}

\begin{equation}\label{A0tan}
A_{0}(\Phi) =  \frac{a_{0}}{\epsilon} \mu\left[  \left( \frac{\Phi}{\Phi_{0}} \right)^{2q}\right] 
\end{equation}
\noindent where $A_{0~max}$ is the maximum value which we allow $A_{0}$ to take $\approx 100 a_{0}$, $\Phi_{0}$ is a scale potential analogous to the MOND scale acceleration with units of m$^{2}$s$^{-2}$ and $q$ is a dimensionless parameter which controls the slope of $A_{0}(\Phi)$. We define $\epsilon$ to be $\epsilon = a_{0}/A_{\rm 0~max}$. Equation \ref{A0tan} says that when the potential is high ($\Phi >> \Phi_{0}$), $A_{0}(\Phi) = A_{0~max}$ and when the potential is low, ($\Phi << \Phi_{0}$), $A_{0}(\Phi) = a_{0}$. This is analogous to the MOND interpolation function, $\mu(x)$.  In the work of \cite{HodsonEMOND}, the parameter choice of $q=2$ was used. In this work, we will also show results for $q=1$. The change in choice of $q$ warrants a change of scale potential, $\Phi_{0}$, as well. For $q=2$, the scale potential is unchanged from \cite{HodsonEMOND} with magnitude $\Phi_{0} \approx -2700000^{2}$ m$^{2}$s$^{-2}$ . For $q=1$, the scale potential is empirically chosen to be $\Phi_{0} \approx -3800000^{2}$ m$^{2}$s$^{-2}$. Therefore given a boundary potential, we can solve Eqn \ref{EMONDPoiss2} and determine the EMOND predicted acceleration profile and hence EMOND predicted dynamical mass.

\section{Modelling The Coma Cluster}\label{Coma}

The first step to modelling the Coma cluster UDGs is to build a model of the Coma cluster itself. We adopt the model of \cite{ComaModel} which has an intra-cluster gas component and cluster galaxy component. There is also a dark matter component in standard gravity, which we can compare with the effective phantom halo predicted by EMOND. 

The gas is modelled via a $\beta$ density profile for which the expression for enclosed mass is,
\begin{equation}
M_{g}(r) = \frac{4}{3} \pi n_{0} (m_{e} + \gamma m_{p}) r^{3} F_{3/2,\beta}\left(\frac{r^{2}}{r_{c}^{2}}\right)
\end{equation}
\noindent where $n_{0}$ is the central electron number density of the emitting X-ray gas in the cluster, $\beta$ is a dimensionless parameter, $r_{c}$ is a scale length of the gas density, $\gamma$ is a parameter which converts the electron number density into a mass density and $F_{\alpha,\beta}(x) \equiv {}_{2}F_{1}\left( 3-\alpha, (3-\alpha)\beta);4-\alpha; -x\right),$ where ${}_{2}F_{1}$ is a hyper-geometric function. 

The galaxies were modelled via,
\begin{equation}
M_{Gal}(r) = 4 \pi L_{\star} \Upsilon r_{s}^{3} \left[ \log\left( \frac{r+ r_{s}}{r_{s}} \right)-\frac{r}{r+r_{s}} \right]
\end{equation}
\noindent where $r_{s}$ is a scale radius, $L_{\star}$ is a luminosity normalisation constant and $\Upsilon$ is a mass-to-light ratio.

Finally, in the work of \cite{ComaModel} the dark matter mass was modelled via,

\begin{equation}
M_{DM}(r) = M_{v} \left( \frac{r}{r_{v}} \right)^{3-\alpha} \frac{F_{\alpha,1}(c r/r_{v})}{F_{\alpha,1}(c)}
\end{equation}
\noindent where $M_{v}$ is the virial mass, $r_{v}$ is the virial radius, $c$ is the concentration and $\alpha$ is a dimensionless parameter.

\begin{figure}
\includegraphics[scale=0.7]{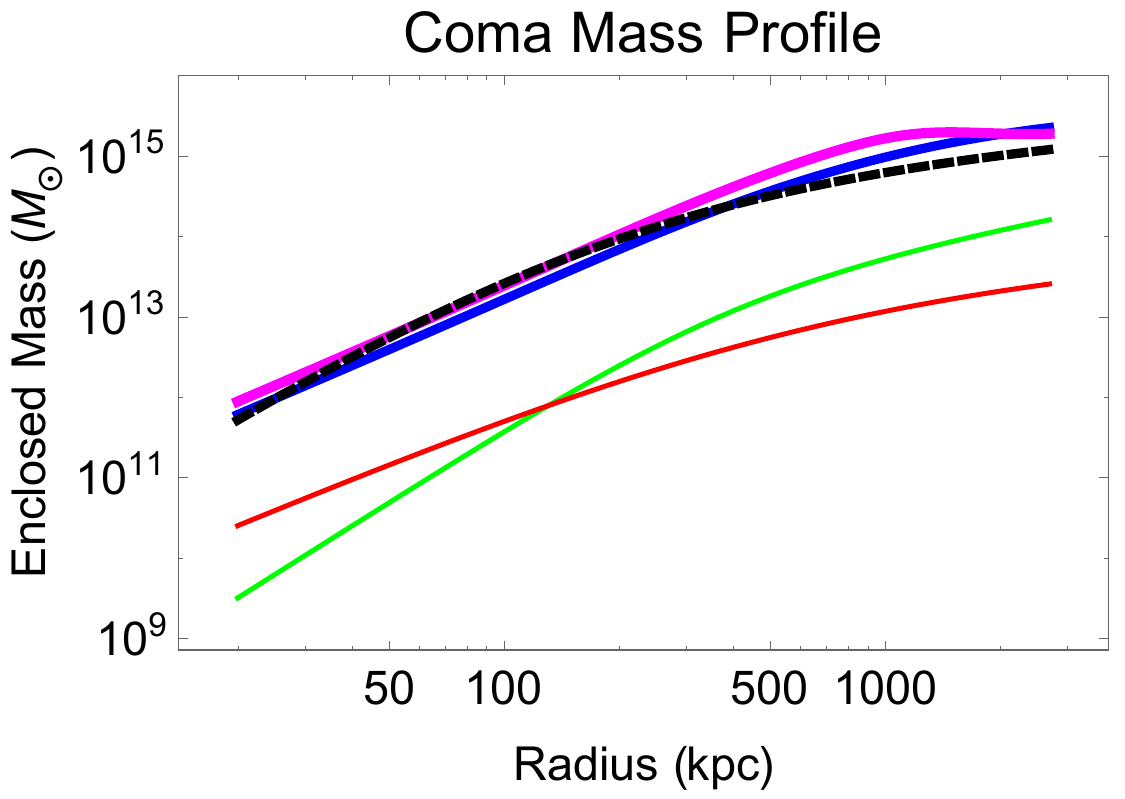}
\caption{Model of the Coma cluster which we adopt from \cite{ComaModel}. Green line shows the contribution from the intra-cluster gas, red thin line is the contribution from the { stars}. Using these, we can calculate the EMOND predicted dynamical mass from Equation \ref{EMONDMass} which is the solid blue line for $q=1$ model and solid magenta line for $q=2$ model (see Eqn \ref{A0tan}). We also plot the dark matter profile from \cite{ComaModel} (black dashed line) for comparison. We see that our EMOND mass matches the dark matter mass very well. For this, we assume an EMOND boundary potential at the virial radius $\Phi(r_{v}) = -2.5 \times 10^{12}$ m$^{2}$ s$^{-2}$. }
\label{ComaMass}
\end{figure}

We plot the mass components of the Coma cluster as a function of radius in Figure \ref{ComaMass}, mimicking the top panel of Figure 8 in \cite{ComaModel}. We over-plot the EMOND predicted mass profile (blue solid line) determined by first solving Equation \ref{EMONDPoiss2}{ for EMOND gravity, which we will call $\nabla \Phi_{EMOND}$ }, and then calculating the effective EMOND mass via,

\begin{equation}\label{EMONDMass}
M_{EMOND}(r) = \frac{r^{2} \nabla \Phi_{EMOND}}{G}.
\end{equation}
{  where the Newtonian gravity used to determine $\nabla \Phi_{EMOND}$ is}
\begin{equation}
\nabla \Phi_{N} = \frac{G (M_{g}(r) + M_{Gal}(r))}{r^{2}}.
\end{equation}

To make the plot, we empirically take a value of the EMOND gravitational potential at the virial radius to be $\Phi(r_{v}) = -2.5\times10^{12}$ m$^{2}$s$^{-2}$\footnote{ The task of EMOND would be to determine the boundary potential from cosmological constraints. Due to the lack of a consistent cosmology we are, at this stage, limited to empirically fitting. Future work on EMOND can determine whether our empirical fit is acceptable.}. We can see that the dark matter dominates over the gas and galaxy contributions. 

The plot also shows that the EMOND predicted mass seems to match the dark matter profile to exceptional accuracy. This is a very good result for the EMOND paradigm. In the previous EMOND work, \cite{HodsonEMOND} found that EMOND had mixed success in describing the clusters. The work in question used a different baryonic mass profile for both the gas and the galaxies. This result suggests that the EMOND modelling of \cite{HodsonEMOND} might be improved by invoking a different functional form for the baryonic mass profile.

\begin{figure}
\includegraphics[scale=0.6]{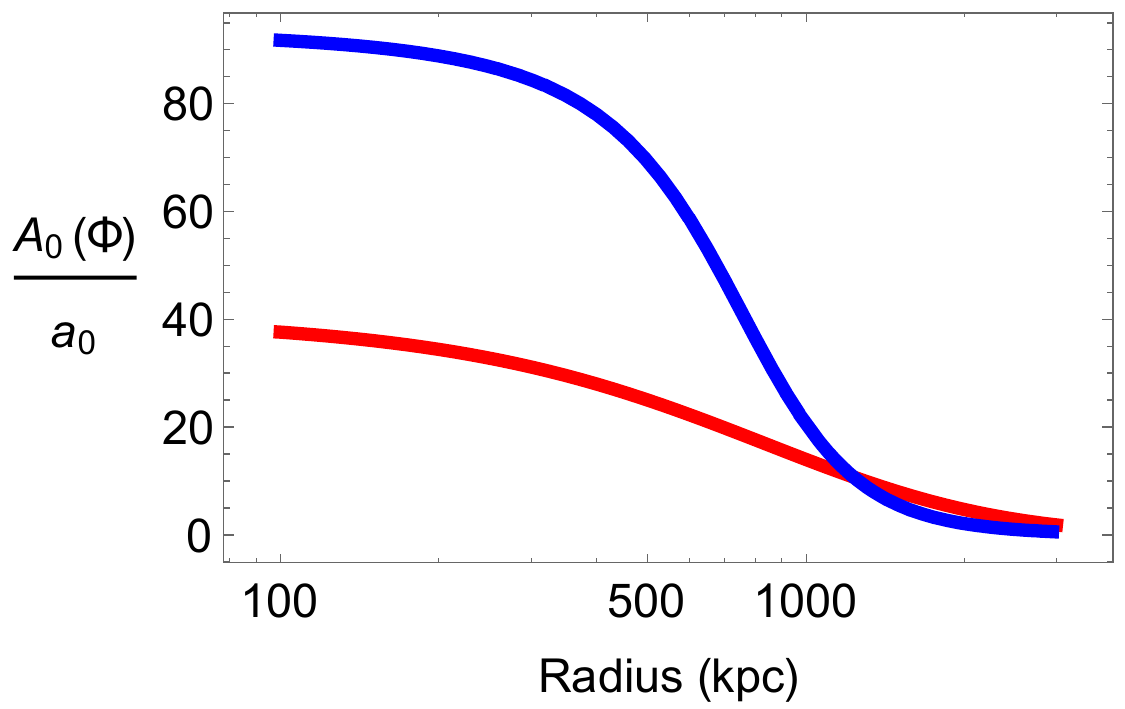}
\caption{Profile of the EMOND calculated $A_{0}(\Phi)/a_{0}$ as a function of cluster radius. Blue dashed line is the $q=2$ model and red solid line is $q=1$ model. The $q=1$ model produces a shallower transition from high to low $A_{0}(\Phi)$ and a smaller in magnitude of $A_{0}(\Phi)$ than the $q=2$ model (see Eqn \ref{A0tan}). We only show radii $> 100$ kpc as this is the important range for the UDGs.}
\label{ComaA0}
\end{figure}

Now that we have derived the EMOND mass profile of the Coma cluster, we can make a plot of $A_{0}(\Phi)$ vs radius to show how the EMOND acceleration scale varies in the cluster environment. We show this in Figure \ref{ComaA0}.

From Fig~\ref{ComaA0}, it is clear that the $q=1$ model creates a gentler transition of $A_{0}(\Phi)$ from the outside of the cluster to the centre than the $q=2$ model. It is also clear that the magnitude of $A_{0}(\Phi)$ which the $q=1$ model predicts is much lower than the $q=2$ model. This is due in part from the gentler transition, but mainly the choice of $\Phi_{0}$, which stops $A_{0}(\Phi)$ reaching $A_{0~max}$. We will see the effect of this in the following sections.

\section{UDG Properties}\label{UDG}

{ A note on conventions. Throughout the following sections we will be referring to several mass quantities which we should  define clearly. The ``dynamical" mass of the UDGs is the inferred mass from dynamics, thus is the total mass of the system. In a $\Lambda$CDM context this would be the mass of the stars + the mass of the dark matter halo. In MOND/EMOND, this would be the baryons + phantom dark matter. The Newtonian mass is the total mass of the baryons, in this case stars. }

When we model the UDGs in EMOND, we will take the dynamical mass of the UDG at the effective radius and determine the Newtonian stellar mass at that radius using the EMOND recipe. We will then compare the Newtonian mass to the estimated stellar mass of these galaxies from their colour and mass-to-light ratio. Therefore, we need to determine both the dynamical mass and stellar mass for these systems. To do this, we follow the techniques used in \cite{UDGFM} and \cite{FM3} for the dynamical mass and \cite{UDGDF} for the stellar mass. We outline the techniques used in these works below.

\subsection{Dynamical Mass}\label{DynamicalMassSec}

The dynamical mass of the UDGs is determined from the velocity dispersion and effective radius via the formula \citep{wolf} (also see  Eqn 1 in \cite{UDGComaObject}),

\begin{equation}\label{MWolf}
M_{\rm dyn}\left|\right._{r_{\rm s}={4 \over 3} r_e} \approx 3  \sigma^2  r_{\rm s}/G = 9.3 \times 10^{5} \sigma^{2} r_{\rm e}
\end{equation}
where $M_{\rm dyn}(<r_{\rm s}) $ is the total enclosed dynamical mass at the spherical half-mass radius $r_{\rm s} \approx {4 \over 3} r_e$, where the $r_{e}$ is the usual effective radius, i.e., the projected circularised half-light radius, $\sigma$ is the velocity dispersion in km/s and $r_{e}$ is the effective 2D radius in kpc. The effective radius was determined and corrected for ellipticity for 46 UDGs within the Coma cluster\footnote{The sample actually has 47 objects, but one object has incomplete data in the table and thus we disregard this entry.} and are given in \cite{UDGDF}. Currently, there is only one UDGs (Dragonfly 44 (DF44)) in the Coma cluster which has a measured value for the velocity dispersion. Therefore, to estimate velocity dispersions for all 46 galaxies in the Coma cluster sample, some assumptions have to be made. 

We have taken a slightly different approach for our study instead of that of  \cite{UDGFM}.
\footnote{In \cite{UDGFM}, the velocity dispersions are determined for the UDGs in the Coma cluster by making use of the fundamental manifold (FM). This relation links effective radius, mean surface brightness within the effective 2D radius and the internal kinematics of the system in question via a nearly power-law-like relation

$\log \Upsilon_{e} = 0.24 \left( \log V \right)^{2} + 0.12 \left( \log I_{e} \right)^{2} - 0.32 \log V - 0.83 \log I_{e} - 0.02 \log \left( V I_{e} \right)+ 1.49.$
where $\Upsilon_{e}$ is the mass-to-light ratio, $I_{e}$ is the mean surface brightness within $r_{e}$ and $V$ describes the kinematics of the system. This was then solved along with the known relation
$\log r_{e} = 2 \log V - \log I_{e} - \log \Upsilon_{e} - C,$
which is derived from Eqn \ref{MWolf}, to determine $V$ and $\Upsilon_{e}$. It was then assumed in \cite{UDGFM} that $V \approx \sigma$. This value of the velocity dispersion was then corrected via $\log \sigma_{corr} = (\log \sigma - 0.061)/0.833$ to account for a {\it `slight systematic deviation from the expectation'}. Therefore \cite{UDGFM} were able to obtain estimates for the velocity dispersions and thus dynamical masses for the UDGs.}

Using the FM relation from a previous study by \cite{FM3}, there exists a relationship between $I_{e}$, $r_{e}$ and $\sigma$, without having to solve the system equations in \cite{UDGFM},

\begin{equation}\label{zaritsky2}
\log r_{e} = -\alpha^{2} \log^{2}\sigma + (2 + 2 \alpha \beta) \log \sigma + B \log I_{e} + C_{2},
\end{equation}

In this equation, $\alpha$, $\beta$ and $C_{2}$ are constants which are empirically determined, taking values (Equation 8 and Figure 11 from \cite{FM1}) $\alpha^{2} \approx 0.63$, $2 + 2\alpha \beta \approx 3.7$, $B \approx -0.705$ and $C_{2} \approx -2.75$. We can use Eqn \ref{zaritsky2} to find the velocity dispersion analytically using the data given in \cite{UDGDF}. The only other difference between our method and \cite{UDGFM} is that we will not make the correction to the velocity dispersion\footnote{The FM line in Figure 11 of \cite{FM1} seems to align well with the data points, hence we do not make a correction.} and assume, for now, that all the UDGs lie on the FM. We will discuss the implications of this later.

The final discussion point is to convert the data table in \cite{UDGDF} to the correct units for the fundamental manifold equation. The fundamental manifold has a 2D effective radius in units of kpc and a mean surface brightness in units of $L_{\odot}/pc^{2}$. To determine the correct radius, we need to take the radii in column 5 (which is the major axis radius) of the table in \cite{UDGDF} and multiply in by the square root of the axis ratio, given in column 7 of the table. For the surface brightness, we need to use a standard conversion to change the central surface brightness, given in column 4 of the table in \cite{UDGDF} in mags/arcsec$^{2}$, into mean surface brightness within an effective radius in $L_{\odot}/pc^{2}$. This is done by

\begin{equation}\label{SBprofile1}
\log <I_{e}> = -\frac{I_{0} + 1.822 - 0.699 - M_{\odot} - 21.572}{2.5},
\end{equation}
where in this case, $M_{\odot}$ is the solar magnitude in the given band, $<I_{e}>$ is the mean surface brightness within an effective radius in $L_{\odot}/pc^{2}$ and $I_{0}$ is the central surface brightness in mags/arcsec$^{2}$. See Appendix \ref{Appendix1} for derivation of Eqn \ref{SBprofile1}.\footnote{The given formula for converting the surface brightness can be more general depending on the S$\acute{\rm e}$rsic index of the modelling. As \cite{UDGDF} used a S$\acute{\rm e}$rsic value of 1 for all UDGs, the above formula is valid for all the galaxies in our sample. }

Once we apply these conversions, we can use Eqn \ref{zaritsky2} to find the estimated velocity dispersion for each UDG and use Eqn \ref{MWolf} to determine the enclosed mass within the 3D radius.

\subsection{Estimating The Stellar Mass at the Effective Radius}

{ In the following sections we will be outlining how to infer the predicted stellar mass in the UDGs in the EMOND paradigm from the dynamical mass estimate described above. Therefore, test the validity of the EMOND formula, we require the approximate enclosed stellar mass at the effective radius for each UDG in the Coma cluster.} In order to do this we follow the technique used in \cite{UDGDF}. This work takes advantage of the relation between colour and mass-to-light ratio, used in \cite{GAMAStellar} which describes a link between the ({\it g}-{\it i}) colour and the stellar mass-to-light ratio in the {\it i}-band,

\begin{equation}\label{Mstellar}
\log_{10} \left[ M_{\star}{\cancel /M_{\odot}}\right] = 1.15 + 0.7(g-i) - 0.4 M_{i}
\end{equation}
     
\noindent where $M_{i}$ is the absolute magnitude in the {\it i}-band and $M_{\odot}$ is the solar mass, not to be confused with the solar magnitude used previously. From this, we can calculate the stellar mass using only colour and magnitude. The g-band magnitude is given for 46 UDG in the Coma cluster in \cite{UDGDF}. For the sample, the average {\it g}-{\it i} colour is  $<g-i> \approx 0.8 \pm 0.1$. { This is the value we adopt for each UDG}. Therefore the {\it i}-band magnitude can be calculated from the quoted g-band magnitude via $M_{i} \approx M_{g}-0.8$. We therefore have all the necessary quantities to derive a stellar mass for the UDGs. Note, the mass calculated via Eqn \ref{Mstellar} is the total mass. The stellar mass within $r_s$, which is what we are interested in, is half of $M_{\star}$.

\subsection{Distance From Centre of Cluster}\label{UDGDistanceSec}

As we only have the 2D projected map of the Coma cluster and the UDGs, it is not possible to get their exact radii from the centre of the cluster. We can however calculate the minimum radius the UDGs should be from the right ascension and declination of the UDGs, as given in \cite{UDGDF}. If we assume that all the UDGs lie at the same distance as the Coma cluster itself, we can find their minimum distance from,

\begin{equation}
d_{\rm UDG-Coma} \approx  d_{\rm Coma}\theta_{\rm UDG-Coma}
\end{equation}
where $d_{\rm Coma}$ is the distance to the Coma cluster and  $\theta_{\rm UDG-Coma}$ is the angular separation in radians between the UDG and the Coma cluster centre
\begin{figure}
\includegraphics[scale=0.7]{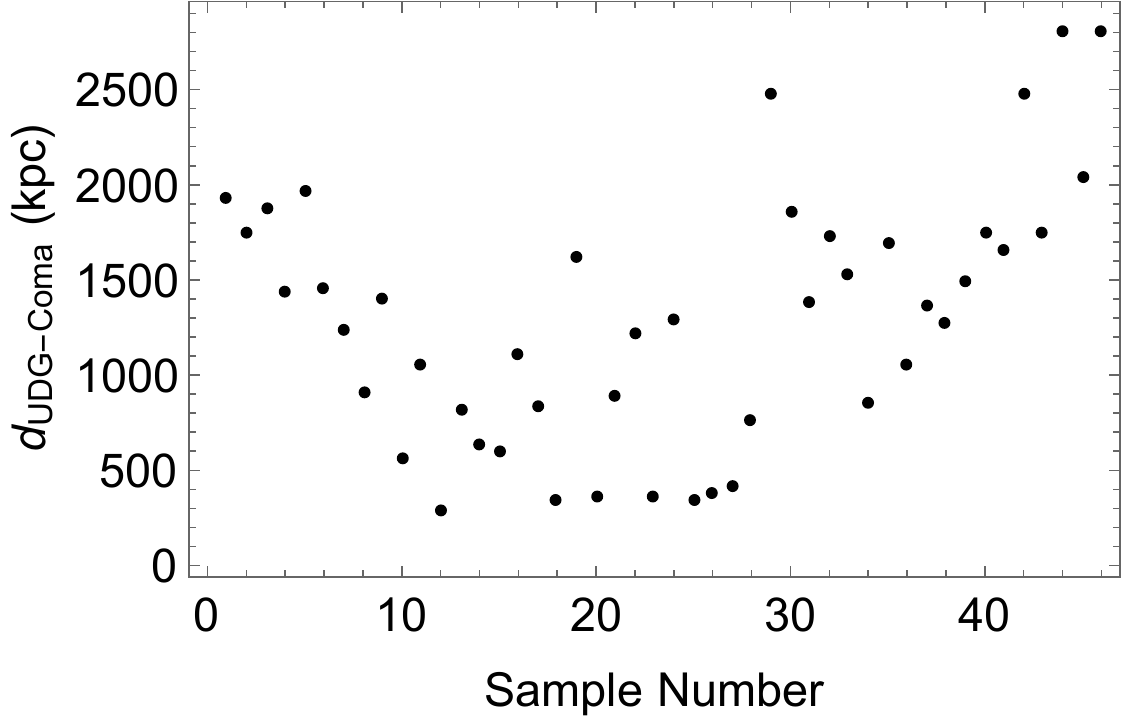}
\caption{The minimum projected distance between the centre of the Coma cluster and the UDGs in kpc. The average distance is approximately 1300 kpc. Note that this is the projected distance and the true 3D distance will be higher than this.}
\label{D_UDG_Coma}
\end{figure}

We can see from Fig \ref{D_UDG_Coma} that the the average distance is approximately 1300 kpc, quite far from the cluster centre, with minimum and maximum values of 296 kpc and 2811 kpc respectively. However, as stated, the actual 3D radii will be on average higher than this.

\section{MOND and EMOND Modelling}\label{MOND}

In this section we describe how the UDGs were modelled in the regular MOND and EMOND paradigms. To do this we take the dynamical mass, derived from the predicted velocity dispersions (Section \ref{DynamicalMassSec}), and substitute the value into the MOND (and EMOND) formula. From this, we can then determine the Newtonian mass which is required to satisfy the MOND equations. Assuming that the galaxy is dominated by stellar mass, we can then compare this Newtonian mass to the stellar mass derived in Section \ref{UDG}. If the MOND paradigm is correct, these two methods should be consistent. All this modelling is conducted under the assumption that the UDGs are spherical\footnote{ The average b/a ratio for the sample is 0.74}.
 
\subsection{MOND}

To begin the MOND modelling, we start by assuming that the UDGs are isolated systems. If they were isolated, we can use the simple spherical MOND relation to model them,

\begin{equation}\label{MOND1}
\nabla \Phi_{Newt} =  \mu\left( \frac{\nabla \Phi_{dyn}}{a_{0}}  \right)\nabla \Phi_{dyn}
\end{equation}
where $\nabla \Phi_{Newt} = G M_{Newt}(r)/r^{2}$ is the Newtonian acceleration and  $\nabla \Phi_{dyn} = G M_{dyn}(r)/r^{2}$ is the dynamical acceleration. As discussed, we can then find the MOND predicted Newtonian mass from the calculated dynamical mass of the UDGs.

However, this is not the correct picture as UDGs are not isolated, they are within the external field of the cluster.  The MOND formula has to be modified to take into consideration the external field of the cluster,\footnote{ The results are found to be nearly the same when we assume $a$ and $g_{ext}$ are orthogonal.}

\begin{equation}\label{MONDPoissExt}
\begin{split}
&\sqrt{(\nabla \Phi_{N})^2 + (\nabla \Phi_{N~ext})^{2}} \approx \\ &\mu \left( \frac{\sqrt{(\nabla \Phi_{dyn})^{2} + (\nabla \Phi_{ext})^{2}}}{a_{0}} \right)\sqrt{(\nabla \Phi_{dyn})^{2} + (\nabla \Phi_{ext})^{2}} 
\end{split}.
\end{equation}

Assuming that the external field is entirely dominated by the Coma cluster, we determine the magnitude of the external field from our model of the Coma cluster in Section \ref{Coma}. The external field used for each UDG is determined from the distance it is from the centre of the cluster, which we calculated in Section \ref{UDGDistanceSec}.

We expect that the external field increases the overall acceleration across the UDGs, pushing the internal dynamics closer to Newtonian as the MOND interpolation function argument is increased. This highlights the tension between the MOND paradigm and the UDG observations. 

\subsection{EMOND}

As we have seen in our Coma cluster EMOND model, the effective value of $a_{0}$ is increased within the cluster. This could raise the dark matter-like effects within the UDGs even with the external field of the cluster dominating the dynamics. This is due to something called the external potential effect. As the UDGs are in the deep potential well of the Coma cluster, under the prediction of the EMOND paradigm, the internal dynamics of the UDGs are affected. The modified version of Equation \ref{MONDPoissExt} for EMOND is,

\begin{equation}\label{EMONDPoissExt}
\begin{split}
&\sqrt{(\nabla \Phi_{N})^2 + (\nabla \Phi_{N~ext})^{2}} \approx \\ &\mu \left( \frac{\sqrt{(\nabla \Phi_{dyn})^{2} + (\nabla \Phi_{ext})^{2}}}{A_{0}\left( \Phi_{dyn} + \Phi_{ext}  \right)} \right)\sqrt{(\nabla \Phi_{dyn})^{2} + (\nabla \Phi_{ext})^{2}} 
\end{split}.
\end{equation}

Making the assumption that $A_{0}(\Phi)$ is approximately constant across the UDGs as they are so small\footnote{Although the gravitational potential of the Coma clusters dominate the UDGs in our model, the gravitational accelerations of the UDGs are still relevant and thus we do not neglect them.}, we can rewrite Equation \ref{EMONDPoissExt} as

\begin{equation}\label{EMONDPoissExt2}
\begin{split}
(\nabla \Phi_{N})^{2} &= \mu \left(  \frac{\sqrt{(\nabla \Phi)^{2} + (\nabla \Phi_{ext})^{2}}}{A_{0}(\Phi_{ext})} \right)^{2}  (\nabla \Phi)^{2} + (\nabla \Phi_{ext})^{2}  \\ &-  \mu \left( \frac{\nabla \Phi_{ext}}{A_{0}(\Phi_{ext})} \right)^{2} \nabla \Phi_{ext}^{2}.
\end{split}
\end{equation}

\noindent where we have eliminated $\nabla \Phi_{N~ext}$ from Eqn \ref{EMONDPoissExt} via $\nabla \Phi_{N~ext} = \mu\left(\frac{\nabla \Phi_{ext}}{A_{0}(\Phi_{ext})}\right)\nabla \Phi_{ext}$.

Equations \ref{MOND1}, \ref{MONDPoissExt} and \ref{EMONDPoissExt2} can then be used to calculate the predicted Newtonian mass of the UDGs given the dynamical mass of the UDGs and the external field and potential, which is derived from the fundamental manifold and the Coma model respectively.

\section{Results}\label{Results}

For our results, we do not perform a rigorous error analysis as there are many sources of error from all the measurements and modelling of the UDGs as well as scatter from the FM and the model of the Coma cluster. We aim to determine simply whether EMOND is a possible explanation for the UDG over-massive dark haloes.

In the following plots we show the ratio of the Newtonian mass, predicted by the MOND and EMOND models, and the stellar mass calculated from the colour. Ideally, this ratio should be 1. If the ratio is less than 1, either the MOND paradigm predicts that there is less mass than is permitted by the stellar mass estimates or the stellar mass estimate is too high. If the ratio is more than 1, the MOND formulation predicts that there is more mass present than is permitted by the stellar mass estimates or the stellar mass estimates are too low. 
 
\begin{figure}
\includegraphics[scale=0.7]{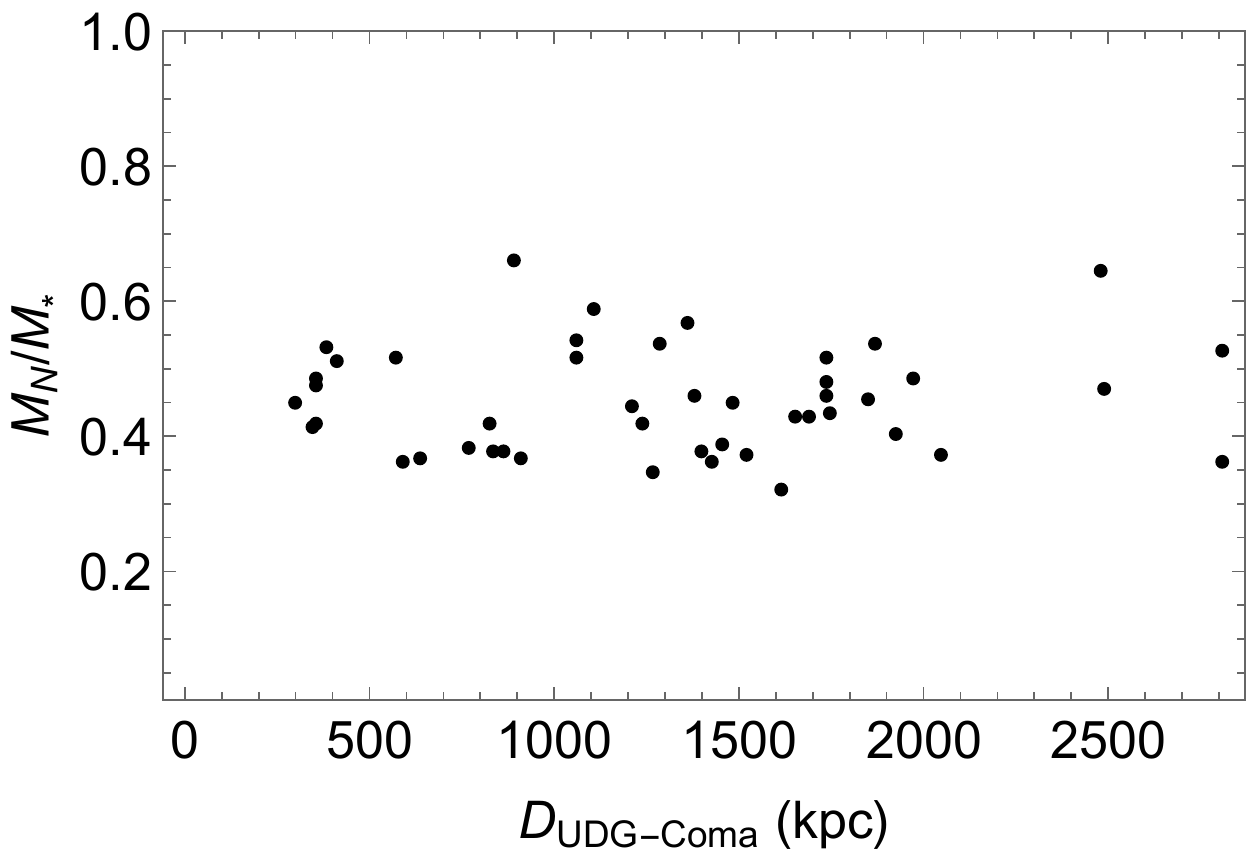}
\caption{Figure showing the ratio of the MOND estimated Newtonian mass to the estimated stellar mass from colour { as a function of the distance to the cluster centre}. No effect from the Coma cluster considered. }
\label{MONDnoExt}
\end{figure}

We begin by showing the result for a MOND model with no effects from the Coma cluster (Figure \ref{MONDnoExt}). We see that for regular MOND the overall trend seems to be that the ratio is less than one by a factor of approximately 2. Therefore, perhaps within the errors, MOND with no external field might be sufficient in explaining the UDG masses.

\begin{figure}
\includegraphics[scale=0.7]{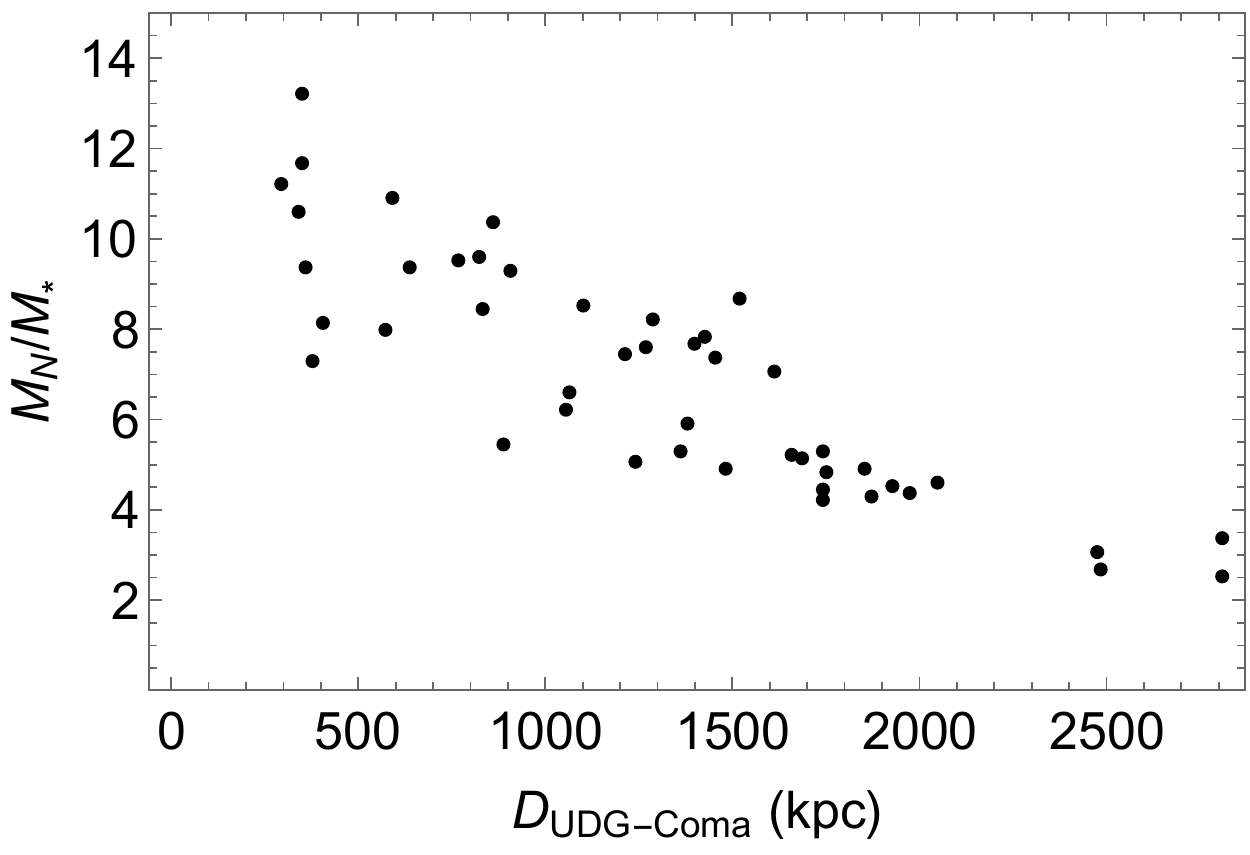}
\caption{Same as Figure \ref{MONDnoExt} except with the inclusion of the external field from the Coma cluster. The MOND predicted mass is much larger than the colour predicted stellar mass.}
\label{MONDExt}
\end{figure}

We next show in Figure \ref{MONDExt} how the external field affects the result. As expected, the cluster boosts the acceleration across the UDGs, increasing the argument in the MOND interpolation function, and thus driving the systems closer to Newtonian. We therefore see that including the external field makes the MOND model worse, the ratio is larger than 1, therefore requiring much more stellar mass than is available according to the colour estimate.

\begin{figure}
\begin{tabular}{c}
\includegraphics[scale=0.7]{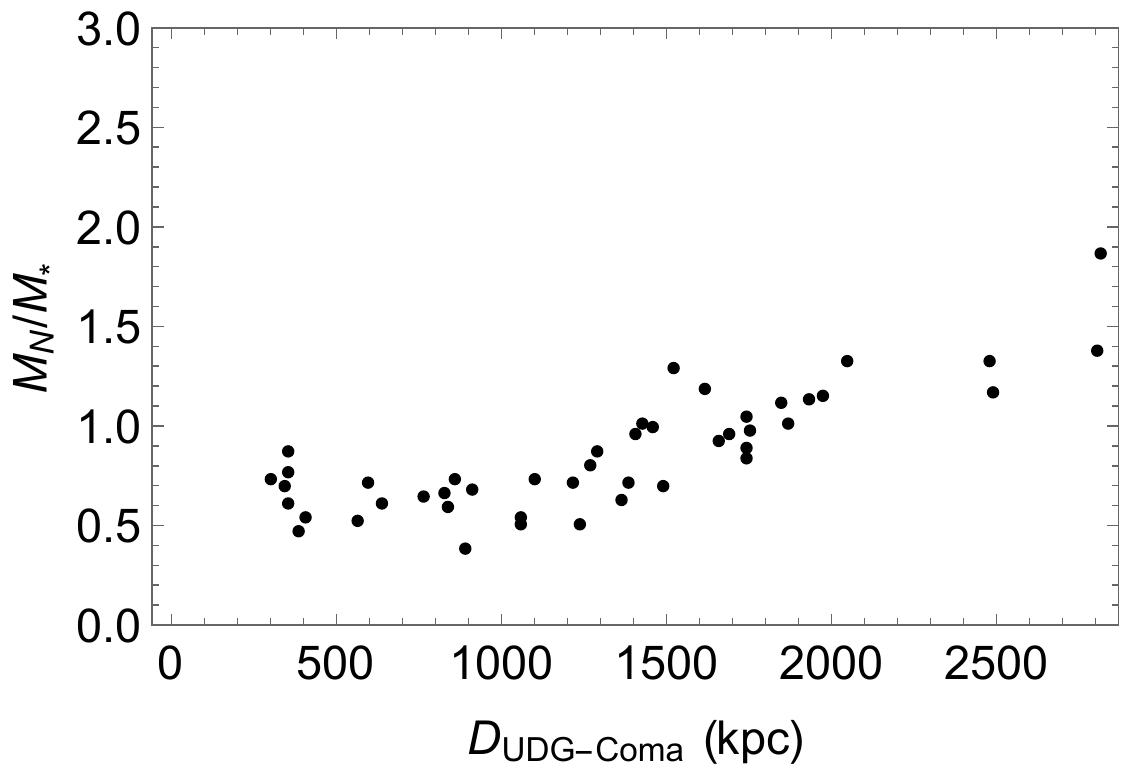}\\
\includegraphics[scale=0.7]{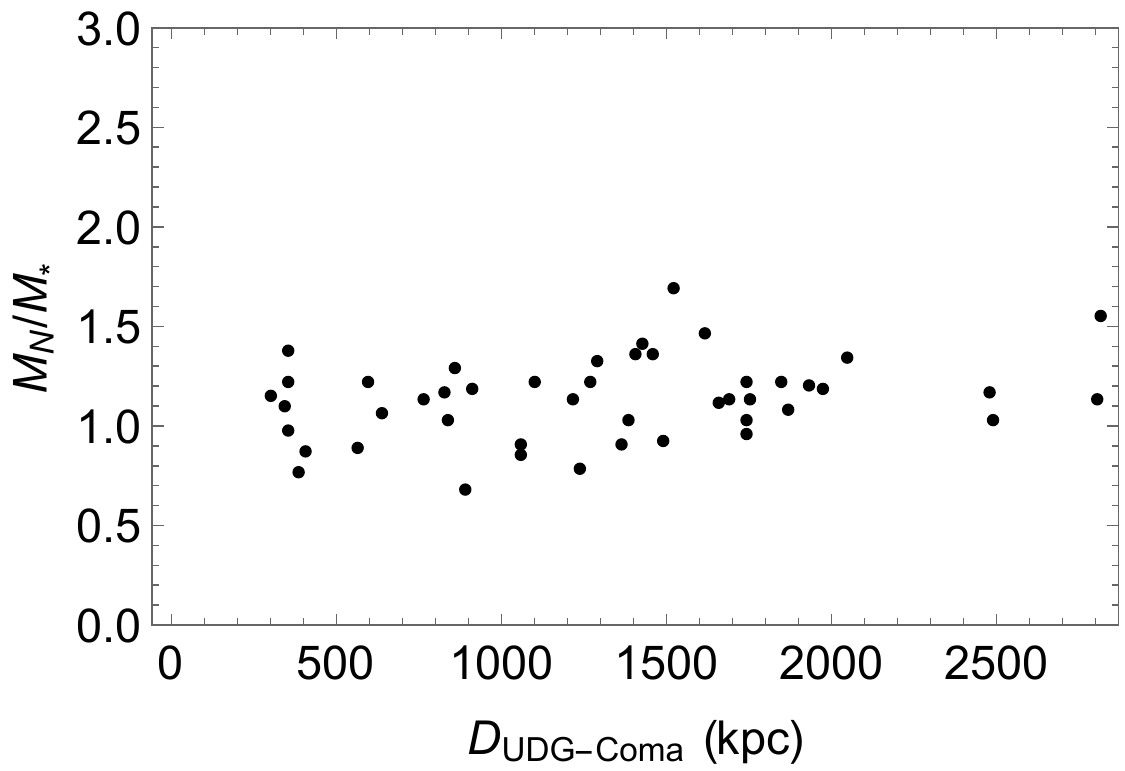}
\end{tabular}
\caption{Same as Figure \ref{MONDExt} except with the EMOND correction to the MOND acceleration scale. Top panel is the $q=2$ model and bottom panel is the $q=1$ model (see Eqn \ref{A0tan}). The EMOND paradigm is predicting a reasonable Newtonian mass for the UDG sample in both models. The $q=2$ model shows that the required mass to light increases with distance, which is an undesirable feature. The $q=1$ model shows that a constant mass-to-light with distance is a good fit to the data, which seems more plausible.}
\label{EMONDExt}
\end{figure}

We then show how the EMOND effect of increasing $a_{0}$ across the UDGs changes the result (Figure \ref{EMONDExt}). We find that the EMOND prediction improves the MOND fit substantially within the expected errors. We also note that in the $q=2$ model (top panel), there seems to be a trend such that the further out the UDG, the higher the predicted Newtonian mass from the EMOND formalism compared to the stellar mass. This is less of an issue with the $q=1$ model, demonstrating that the UDGs provide a stringent constraint in the allowed functional form of $A_{0}(\Phi)$. This might be an indication that rigorous numerical testing and a larger sample of UDGs might find that further refining the EMOND parameters and interpolation function might produce an even better fit. This is beyond the scope of this paper. Another point of note is the fact that the outer UDG values are similar in the MOND and EMOND case. This is due to the EMOND formalism asymptotically tending to MOND in the outer part of the cluster, as desired.      

The above results seem to show that if we take at face value the dynamical mass of the UDGs, stellar mass of the UDGs, EMOND function and the model of the Coma cluster, that EMOND is able to explain the Coma cluster mass profile, and the UDGs within it. The $q=1$ model produces a better fit to the data than the $q=2$ model in terms of how the distance of the UDGs from the centre of the Coma cluster is affected by EMOND.

There will undoubtedly be sources of error within this calculations from spherical symmetry assumptions, scatter around the FM, the error in the Coma cluster mass model etc which will alter the result. The main source of error is most likely the uncertainty in the stellar mass-to-light ratio and the use of the M/L - (g-i) relation.

\begin{figure}
\begin{tabular}{c}
\includegraphics[scale=0.7]{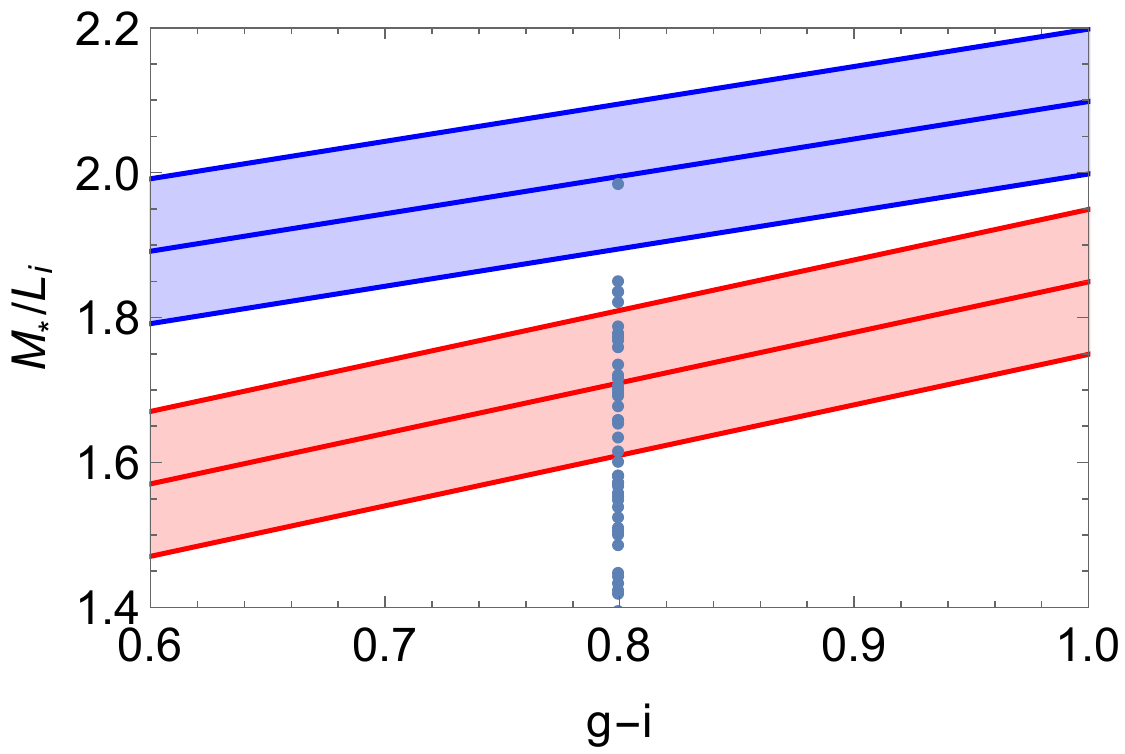}\\
\includegraphics[scale=0.7]{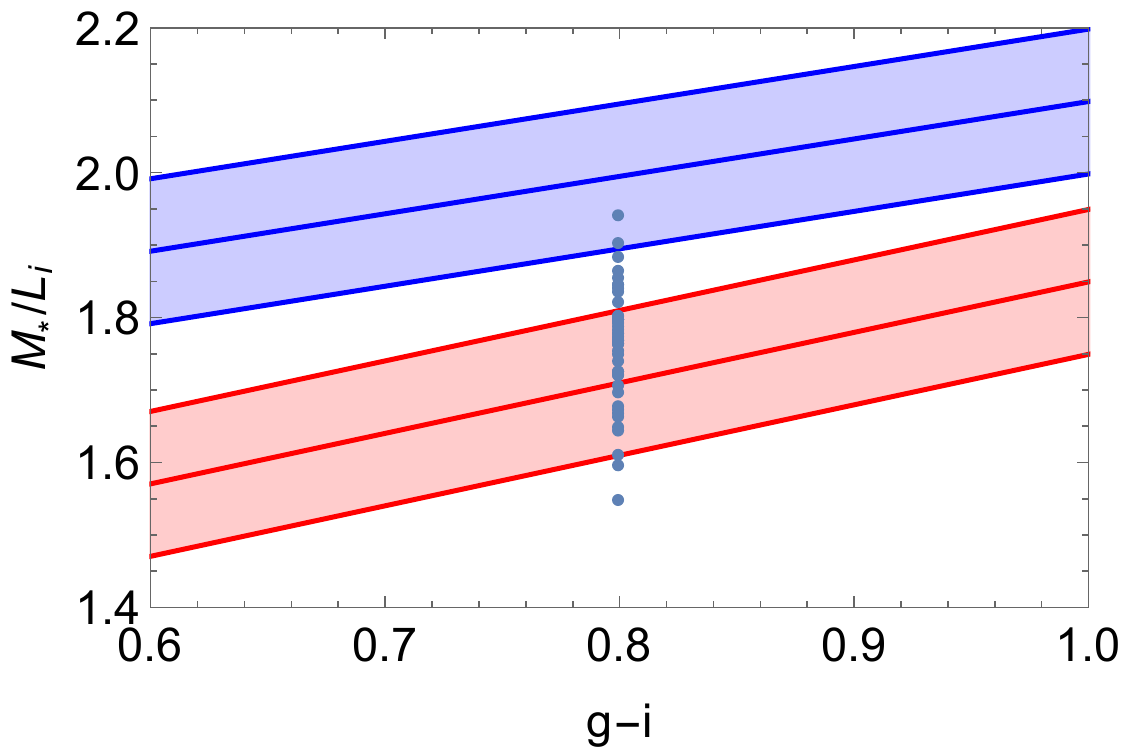}
\end{tabular}
\caption{Plot showing the stellar mass-to-light functions from \cite{GAMAStellar} (red) and \cite{BellStellar} (blue) as a function of g-i colour. We show approximate error bars of 0.1 dex for each case. Top panel shows results for $q=2$ model and bottom panel shows the $q=1$ model (see Eqn \ref{A0tan}). The blue dots show where the UDGs must lie assuming that the EMOND formulation is correct. This shows that it may be possible for most of the UDGs to be explained by EMOND within the range of stellar mass-to-light allowed. The $q=1$ model again shows better results.}
\label{GAMAPlot2}
\end{figure}

To get an idea of the error in the stellar mass-to-light ratio, we recreate Figure 13 from \cite{GAMAStellar} in Fig~\ref{GAMAPlot2} with the stellar mass-to-light used in this work (red band) and the \citep{BellStellar} function (blue band).  There is quite a bit of contention between these two estimates of the M/L ratio. It is possible to reverse engineer the question by assuming that the EMOND formalism is correct and determining the required value of the stellar mass-to-light ratio of each object. For this, we assume that each UDG lies on the <{\it g}-{\it i}> = 0.8 line.  We assume both functions have an approximate error of 0.1 dex (coloured band region for each function), which is reasonable according to the literature \citep{GAMAStellar}. We then determine the required value of the stellar mass to match the EMOND predicted mass and determine where on the mass-to-light plot each UDG lies. We show this in Fig \ref{GAMAPlot2}. We should remind ourselves at this point that the 0.8 value is an average with an error of $\pm 0.1$, therefore there is an extra source of uncertainty.   

Figure \ref{GAMAPlot2} shows that there seems to be a very large scope for error in the mass-to-light of the stars on the UDG, within which, most UDGs in our sample lie. We can therefore conclude that adjusting the stellar mass-to-light ratio can explain the UDGs mass, within the error bars, assuming that the EMOND modelling of the UDGs is valid.

\section{Adjusting the EMOND Formulation}\label{refine}

The above results show that the $q=1$ model fits the UDGs better than the $q=2$ model used in \cite{HodsonEMOND}. Therefore, for completeness, we should redo the analysis of \cite{HodsonEMOND} to check the $q=1$ model is consistent with the cluster sample of \cite{sample}. To do this, we will briefly review the \cite{HodsonEMOND} work and recreate their Figures  17-22 with the updated function for $A_{0}(\Phi)$.

One method of testing modified gravity theories is by comparing the estimated mass, derived from Newtonian dynamics and the mass calculated by assuming the inter-cluster gas is in hydrostatic equilibrium, which we call the dynamical mass. We have discussed how to find the Poisson predicted mass in the above sections. The dynamical mass is determined by solving the equation of hydrostatic equilibrium,

\begin{equation}\label{DynamicalMass}
M_{dyn}(r) = -\frac{kT(r)r}{G wm_{p}}\left[ \frac{d \ln \rho_{g}(r)}{d \ln r} + \frac{d \ln T(r)}{d \ln r} \right]
\end{equation}

where $\rho_{g}(r)$ is the density of the gas, $T(r)$ is the temperature of the gas, $k$ is the Boltzmann constant, $m_{p}$ is the proton mass and $w$ is the mean molecular weight. Therefore, for a given gas density and temperature, the dynamical mass can be calculated. 

The last aspect to discuss is the determining of the boundary potential used to solve the Poisson Equations for each cluster. To get an estimate, \cite{HodsonEMOND} used the analytical best fit NFW profiles for each cluster and assumed that $\Phi(r_{\rm out}) \approx \Phi_{NFW}(r_{\rm out})$ where $r_{\rm out}$ was defines as some boundary outside the cluster. They then showed the range of solutions from  $\Phi(r_{\rm out}) = (0.5 - 1.5) \times \Phi_{NFW}(r_{\rm out})$ to get an idea of how changing the boundary potential affects the result. { Here, to be consistent, we set each boundary potential to take the same value used for the Coma cluster. We also have modelled the galaxies for each cluster to have a similar mass profile as the Coma cluster in the central regions. { We note that each cluster will in practice have a different baryonic profile for the galaxies within the cluster.}  We also show the boundary potential for $\Phi(r_{v}) = (0.9 - 1.1) \times \Phi(r_{v})$ in contrast to the previous work.} Better fits might be possible by numerically playing with this value, not addressed here\footnote{Better fits may also be found by using a different galaxy and/or gas model. Again, not addressed here.}.

\begin{figure*}
\begin{tabular}{ccc}
\includegraphics[scale=0.5]{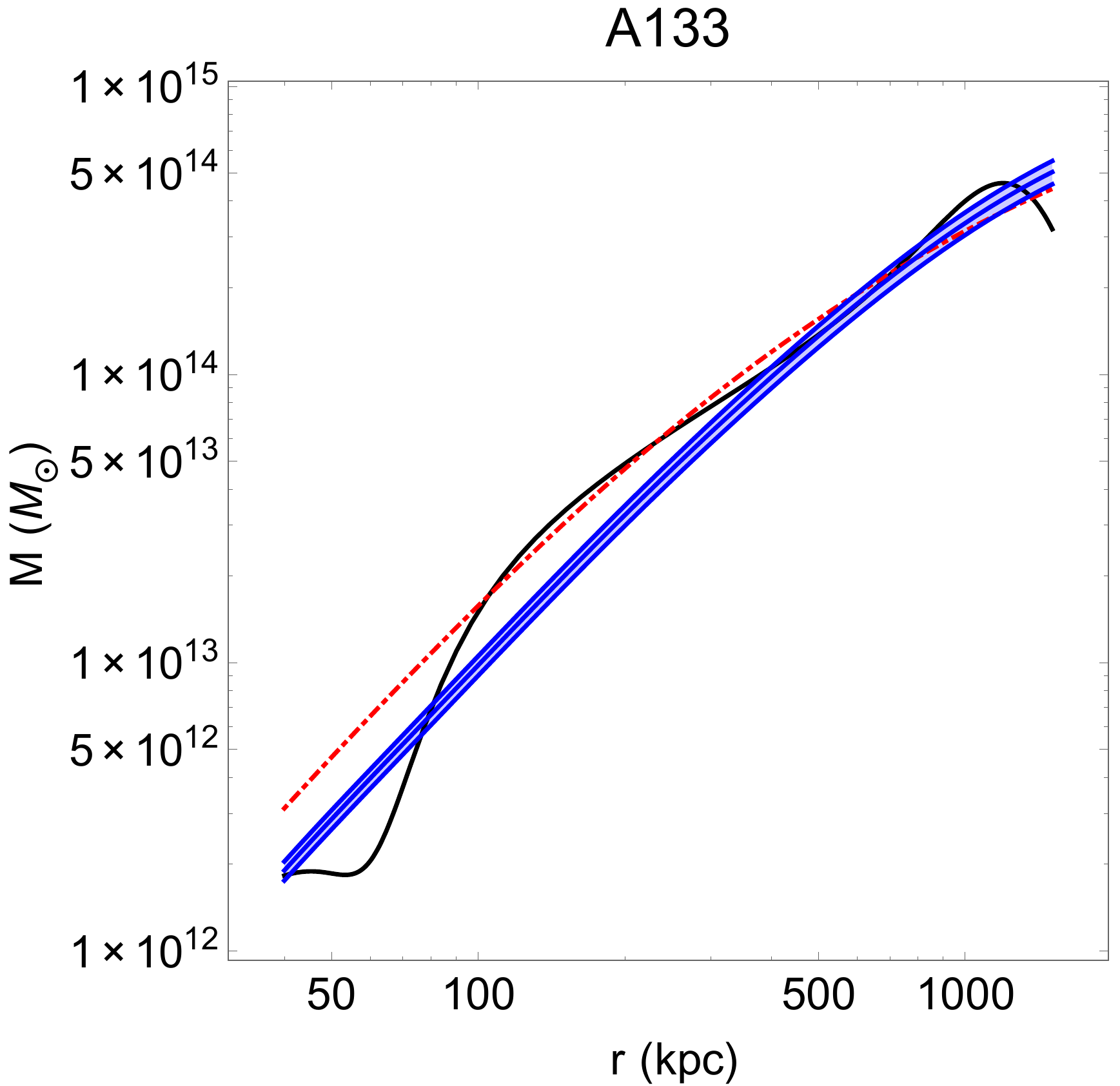} & \includegraphics[scale=0.5]{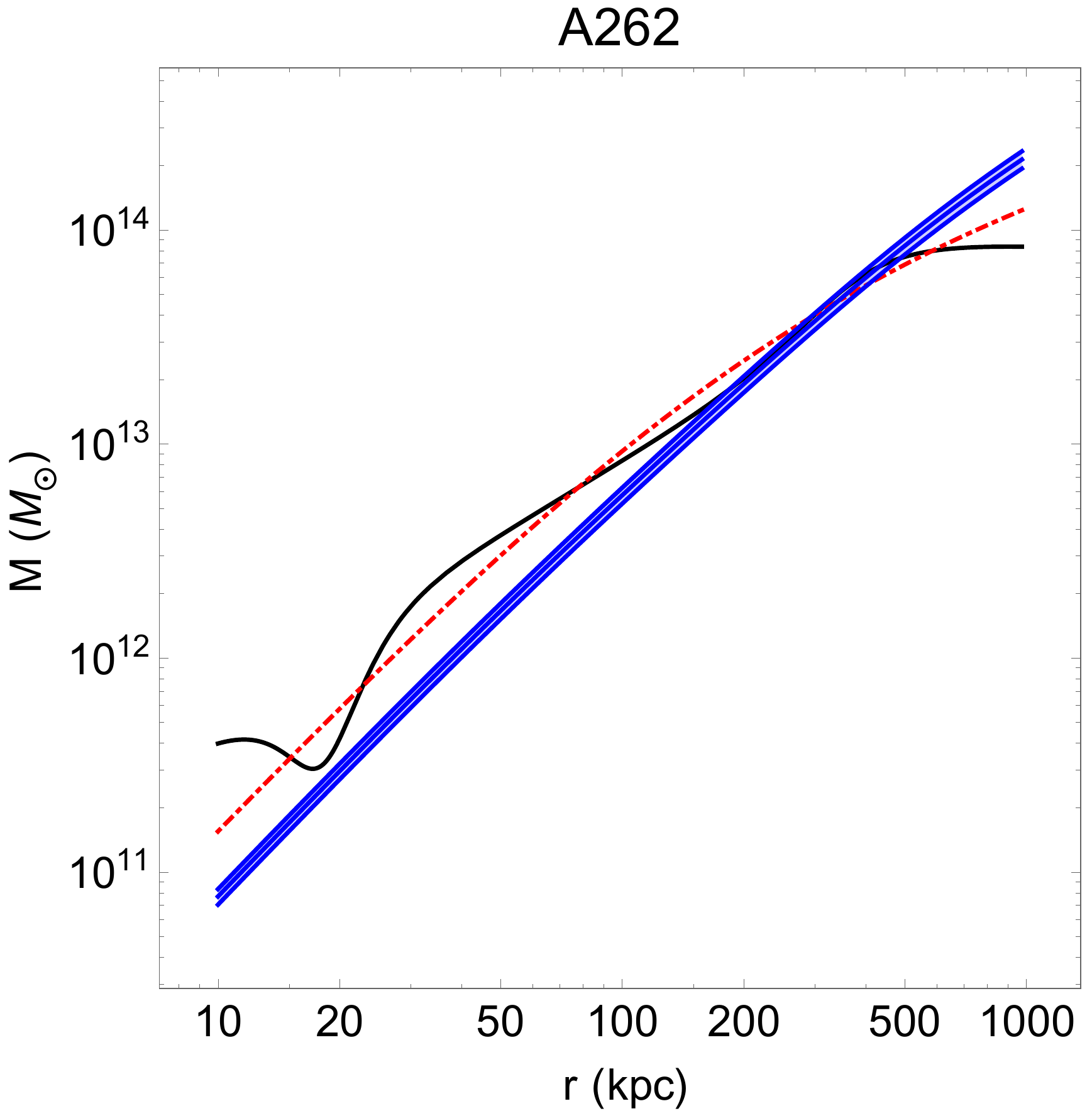}
\end{tabular}
\caption{Plot showing recreated Figs 17-22 from \cite{HodsonEMOND} with the modified $A_{0}(\Phi)$ function found under the UDG constraints. { Red dashed line is the best-fit $\Lambda$CDM model from \cite{sample}, black line is the dynamical mass derived from Eqn \ref{DynamicalMass} and the blue shaded region is the EMOND predicted mass. Here we show clusters A133 and A262.}}
\label{MassPlotEMOND}
\end{figure*}

Therefore, redoing the above steps for the new $A_{0}(\Phi)$ function, we show the updated mass plots for the cluster sample (Figs \ref{MassPlotEMOND} - \ref{MassPlotEMOND6}).

\begin{figure*}
\begin{tabular}{ccc}
\includegraphics[scale=0.5]{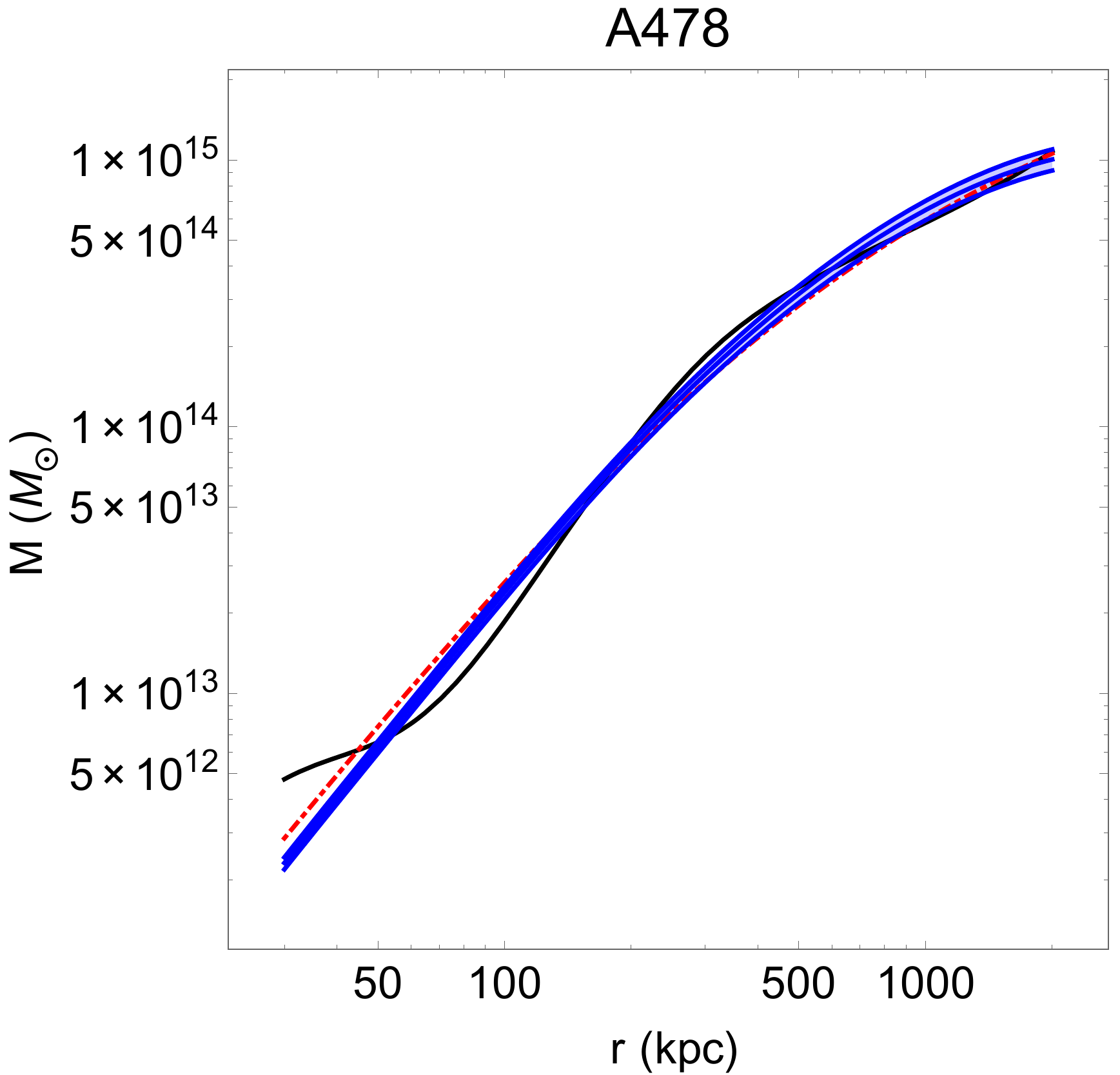} & \includegraphics[scale=0.5]{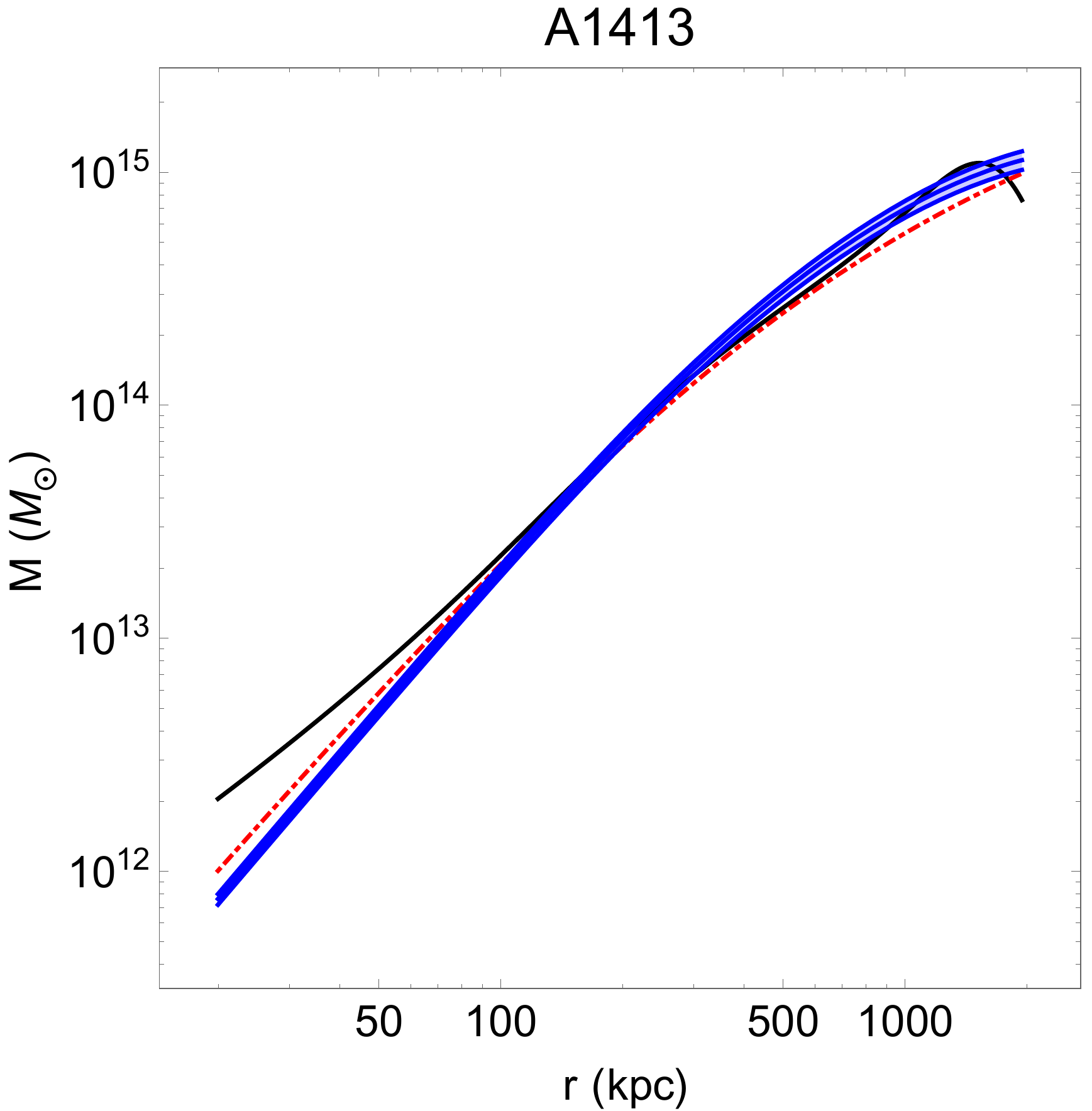}
\end{tabular}
\caption{Same as Figure \ref{MassPlotEMOND} for clusters A478 and A1413.}
\label{}
\end{figure*}

\begin{figure*}
\begin{tabular}{ccc}
\includegraphics[scale=0.5]{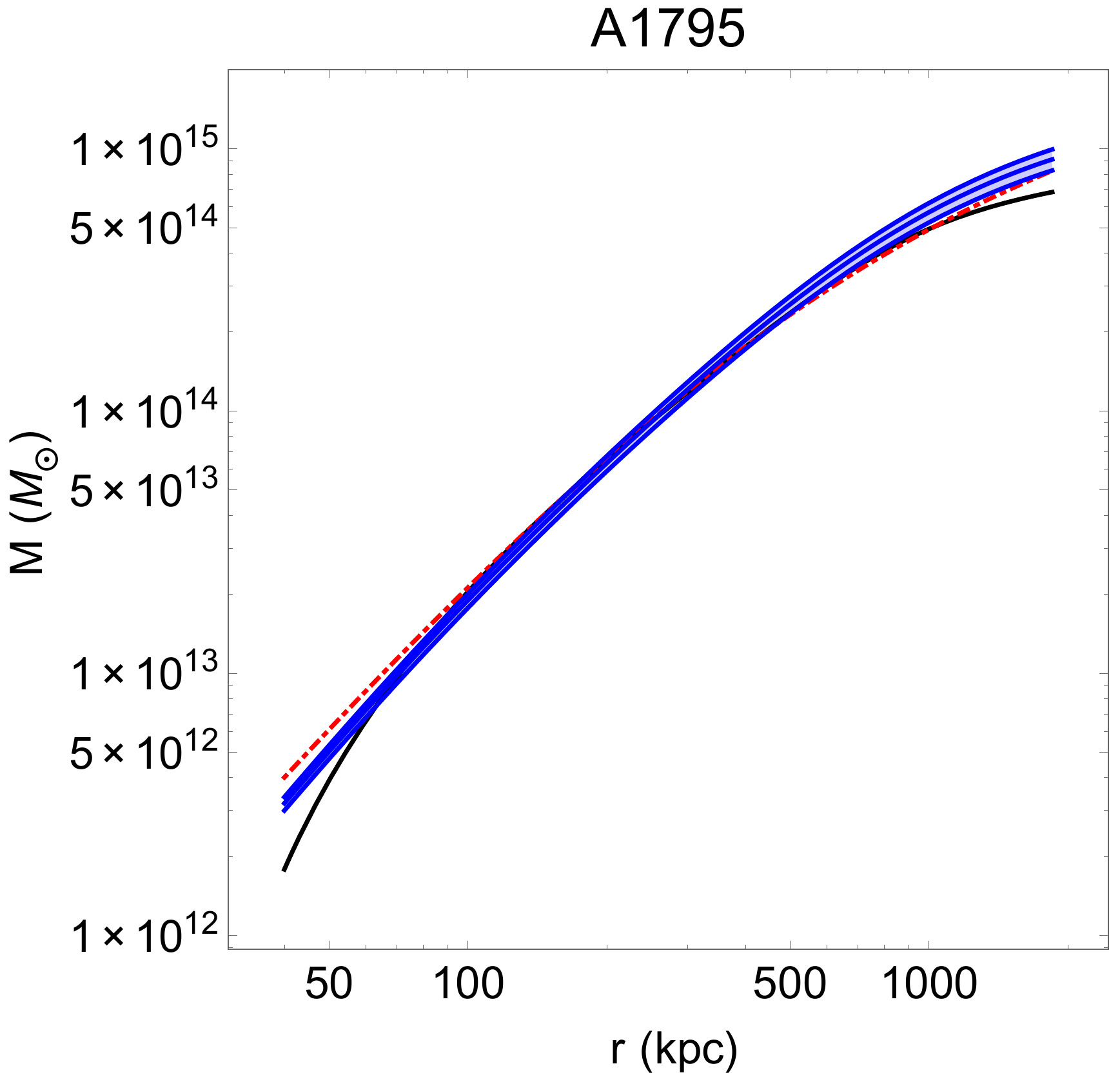} & \includegraphics[scale=0.5]{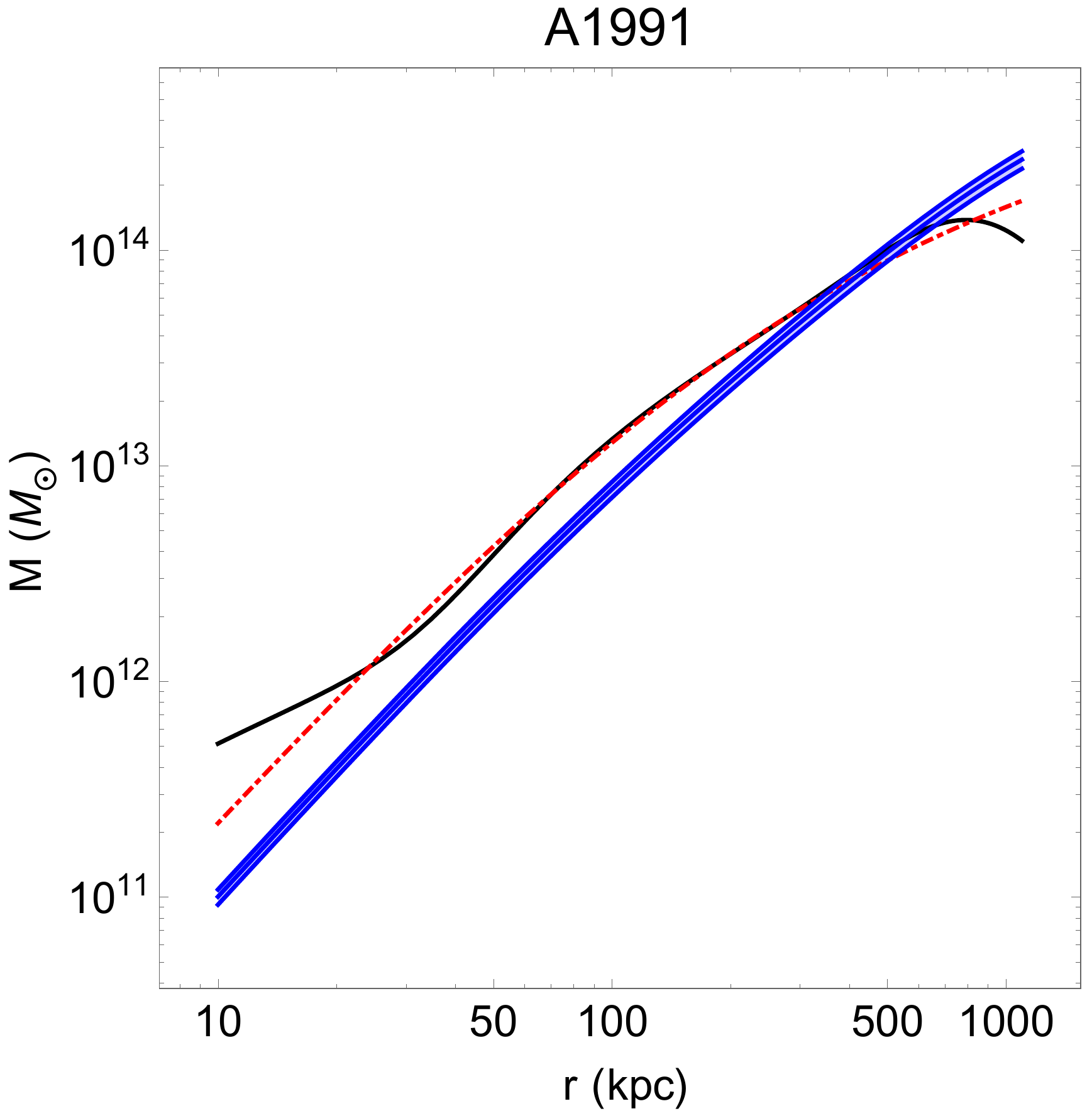}
\end{tabular}
\caption{Same as Figure \ref{MassPlotEMOND} for clusters A1795 and A1991.}
\label{}
\end{figure*}

\begin{figure*}
\begin{tabular}{ccc}
\includegraphics[scale=0.5]{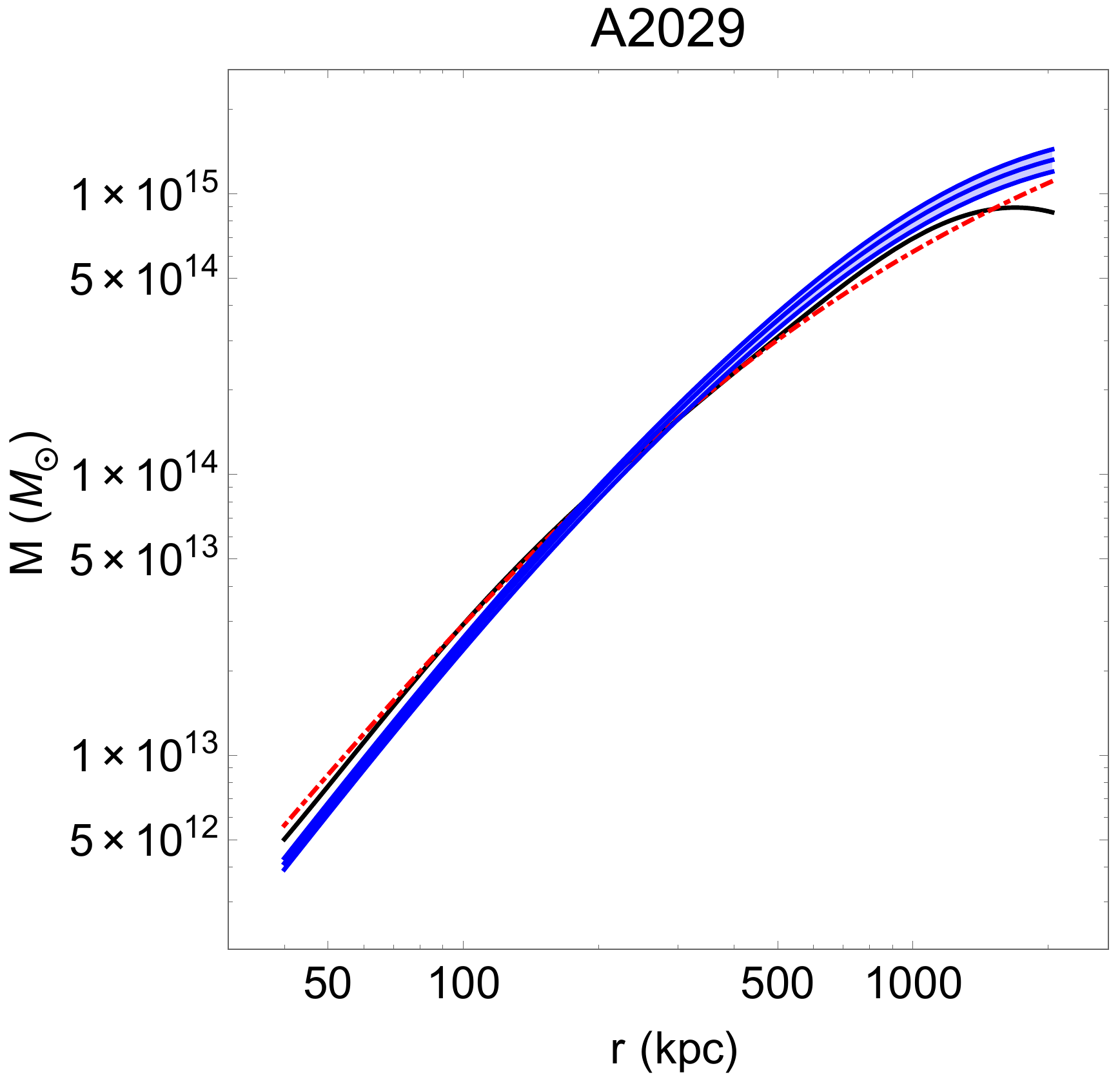} & \includegraphics[scale=0.5]{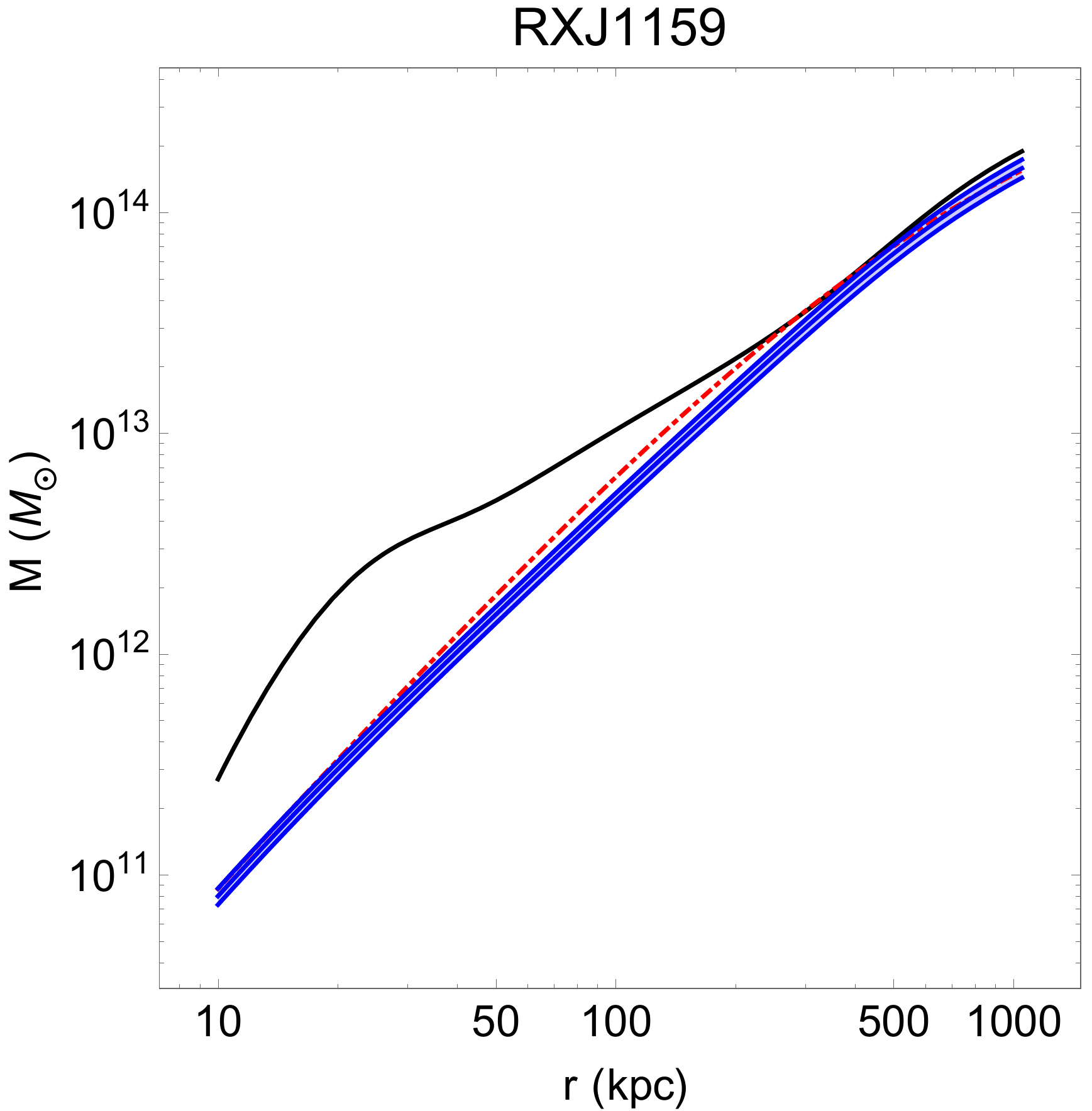}
\end{tabular}
\caption{Same as Figure \ref{MassPlotEMOND} for clusters A2029 and RXJ1159.}
\label{}
\end{figure*}

\begin{figure*}
\begin{tabular}{ccc}
\includegraphics[scale=0.5]{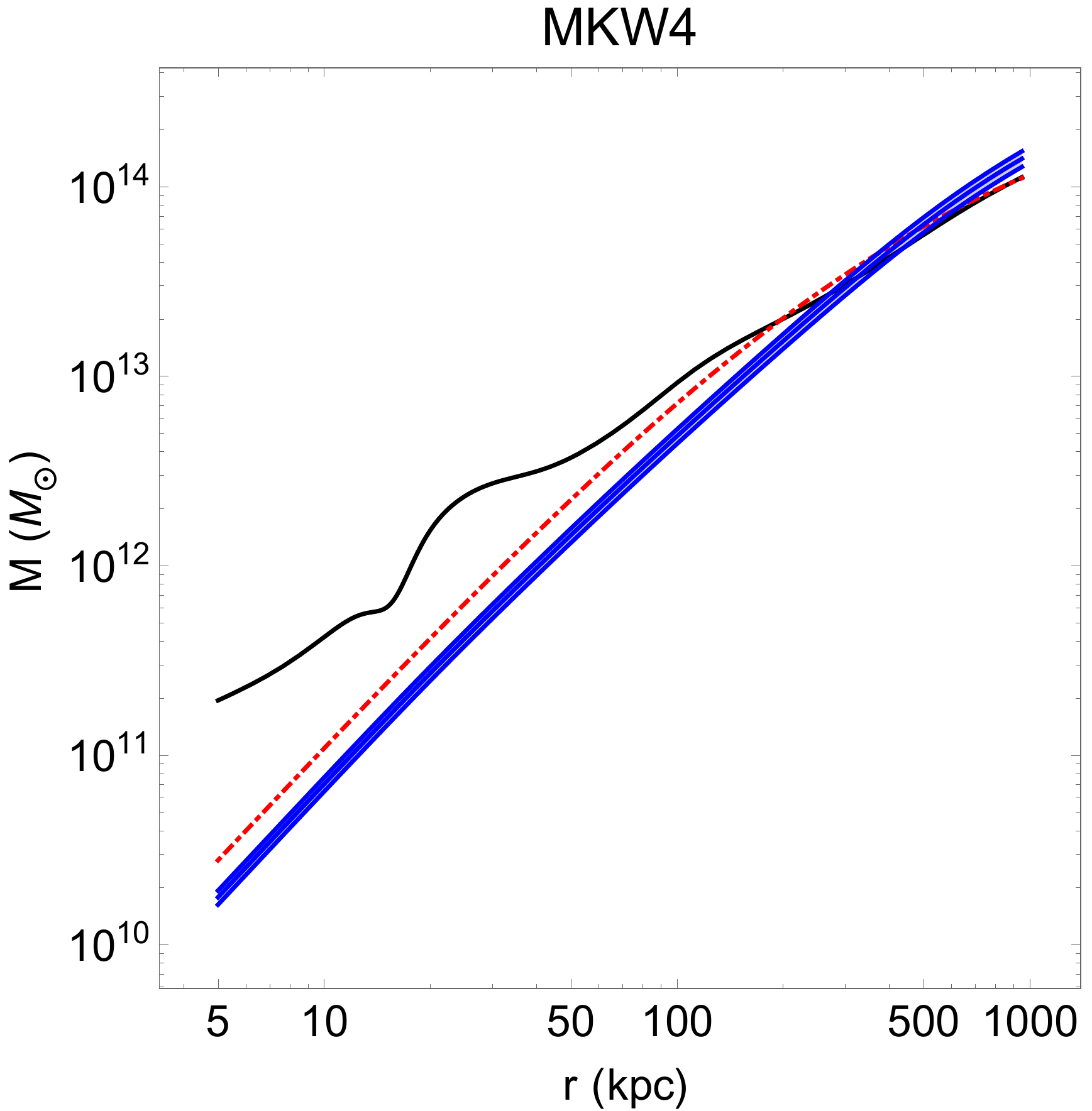} & \includegraphics[scale=0.5]{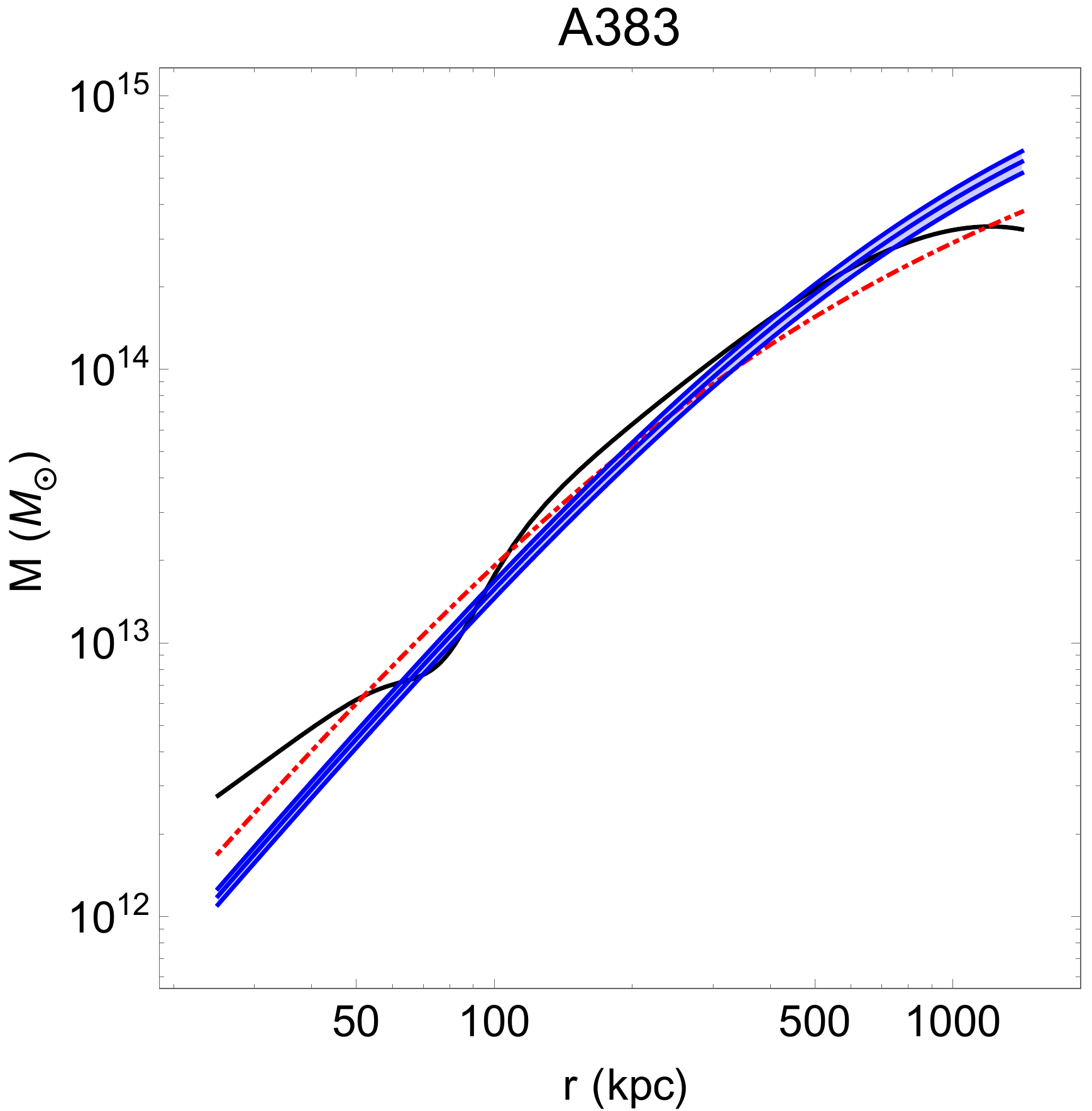}
\end{tabular}
\caption{Same as Figure \ref{MassPlotEMOND} for clusters MKW4 and A383.}
\label{}
\end{figure*}

\begin{figure*}
\begin{tabular}{ccc}
\includegraphics[scale=0.5]{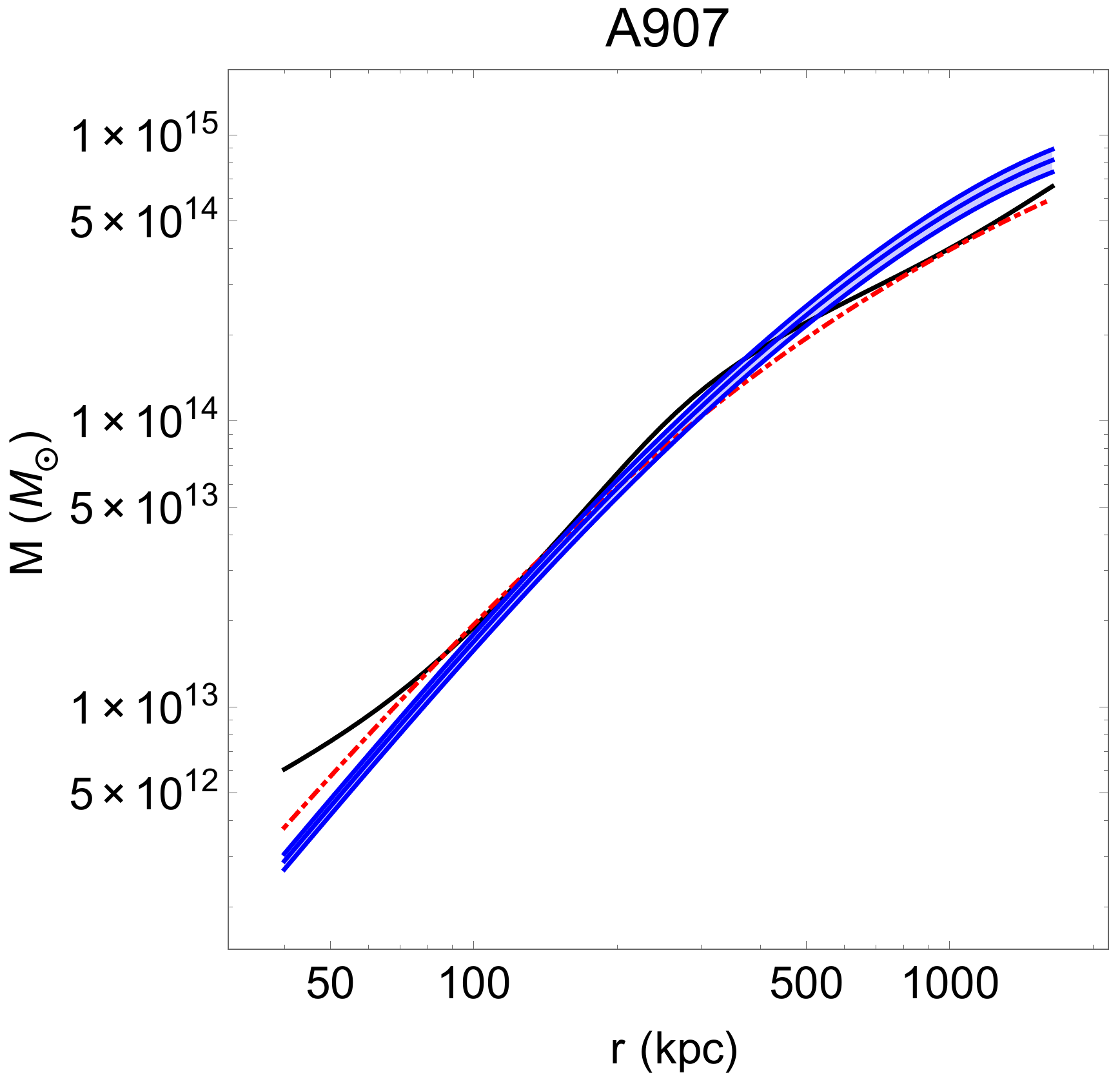} & \includegraphics[scale=0.5]{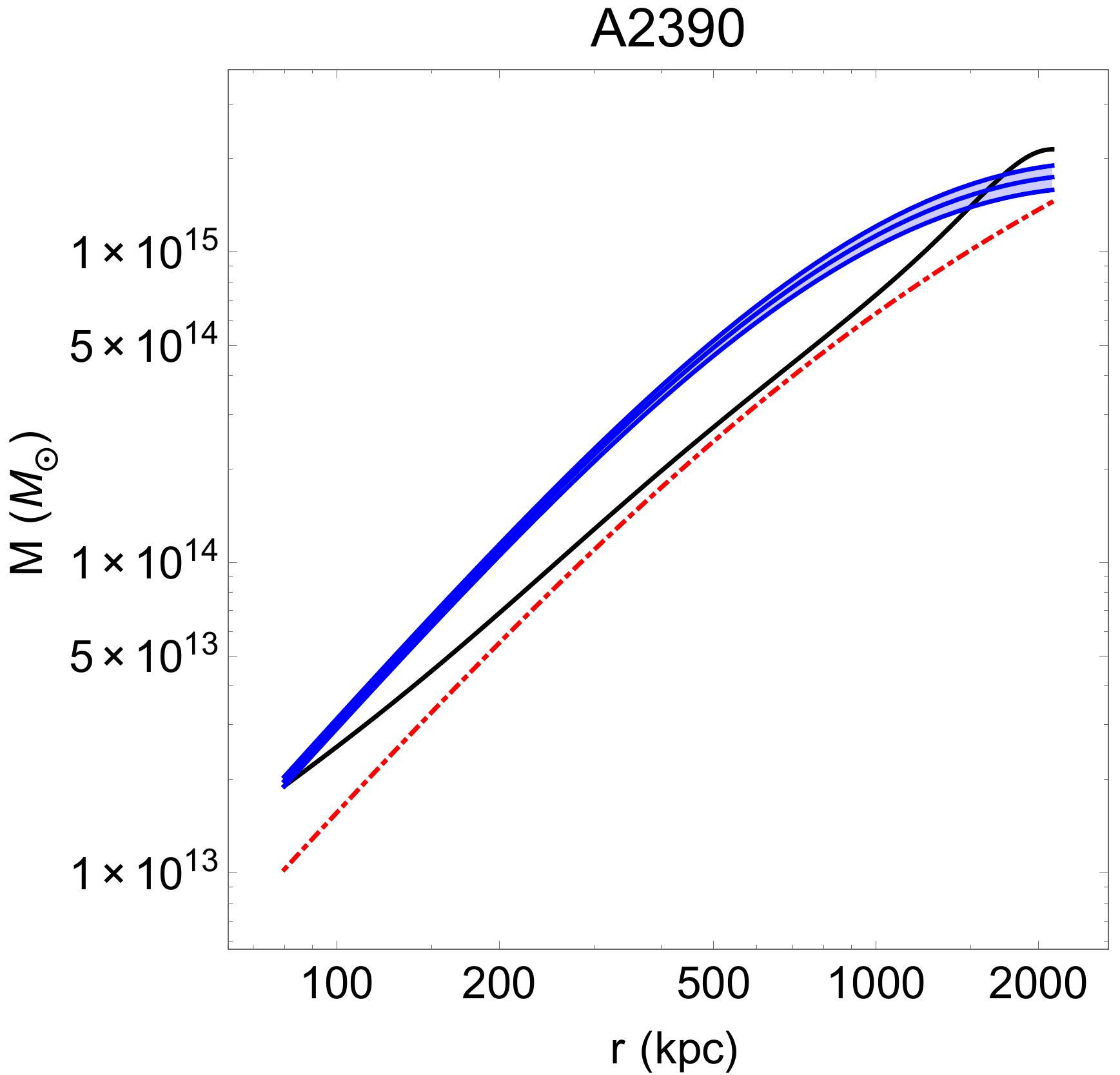}
\end{tabular}
\caption{Same as Figure \ref{MassPlotEMOND} for clusters A907 and A2390.}
\label{MassPlotEMOND6}
\end{figure*}

In these plots, the blue denotes the predicted mass by EMOND with the shaded region showing how the value is affected by different choices of the boundary potential, the red dashed curve is the NFW prediction from \cite{sample} and the black line is the dynamical mass, predicted from hydrostatic equilibrium. The EMOND blue curve should match the black curve, within errors related to modelling assumptions. We can see from Figs \ref{MassPlotEMOND}-\ref{MassPlotEMOND6} that the new form of $A_{0}(\Phi)$ gives a better fit than those found in \cite{HodsonEMOND}.{ It is also possible to see in cases such as RXJ1159 and MKW4 that even though the EMOND predicted dynamical mass is far from the dynamical mass from hydrostatic equilibrium, the EMOND curve is consistent with the best fit NFW curve (red dashed). Therefore we can conclude that changing the form of the interpolation function is still consistent with the previous EMOND work.}

\section{Contention With Observation}\label{contention}

Although our above analysis has shown a consistency between the Coma cluster and UDGs masses under the EMOND paradigm, we have made some rather large assumptions, the main assumption being that the FM can be used to determine the velocity dispersion of the UDGs. If we take the estimate for DF44, which is the only UDG in the sample which has been observed ($\approx$ $47^{+8}_{-6}$ km/s), the FM under-predicts the velocity dispersion by a factor of $\approx$ 2.7 (FM predicted velocity dispersion for DF44 is $\approx$ 17.4 km/s). If we then take the published observed data from \cite{UDGComaObject}, EMOND would predict a Newtonian mass of $\approx$ $7.7 \times 10^{8}$ M$_{\odot}$ and thus the ratio between this EMOND (q=1 model) predicted Newtonian mass and the stellar mass is $\approx$ 7.5 which is quite a substantial difference. This result is improved if the EMOND boundary potential chosen for the Coma cluster is increased. Choosing the boundary to be $3.5 \times 10^{12} \rm m^{2}s^{-2}$, the ratio is reduced to $\approx$ 6. If this potential was chosen, and we took the lowest bound for the velocity dispersion (41 km/s), the  ratio is further decreased to $\approx$ 4.5. This could be further improved by choosing a higher stellar mass-to-light ratio than is used in \cite{UDGComaObject}. However, it must be checked what values for the boundary potential are allowed by the data for Coma. This would require further work, beyond the scope of this paper. 

The reason our result differs from the work of \cite{UDGFM} is that the velocity dispersion is corrected due to there being a discrepancy between the observed velocity dispersion and the FM estimated value (see Figure 1 of \cite{UDGFM}). In our analysis, we used a different form of the FM. Further study as to the source of the discrepancy should be investigated in further work. 

More detailed observations of more UDGs in the Coma cluster will be required to determine whether the over-massive dark halo of DF44 is a statistical outlier  in the sample, or whether the interpretation of the FM used in our work is at a contention with the current observations.

\section{Conclusion}\label{Conclusion}

In this work, we modelled the Coma cluster in the EMOND paradigm and compared the predicted enclosed mass profile to that of a pure Newtonian model. We find that the EMOND result bears an extraordinary resemblance to the DM profile used in \cite{ComaModel}. This is quite a successful result for the EMOND paradigm. The success of this result should warrant further study into EMOND, taking into careful consideration the functional form of the baryonic mass profile and the boundary potential used to solve the Poisson equation. 

We then moved on to make a model of UDGs in EMOND. We used this to determine the predicted Newtonian mass required to satisfy the EMOND formula. We then compared this to the stellar mass predicted by the UDG galaxy colour.

Our model seemed to give consistent values of the EMOND predicted Newtonian mass and the stellar mass derived from colour, within the error bars of the stellar mass-to-light. Further to this, the UDG sample gave a constraint on the exact function of  $A_{0}(\Phi)$. Using a slightly different function to that of \cite{HodsonEMOND} yielded better results. This function was also checked against the earlier work of \cite{HodsonEMOND} yielding not only consistent, but better results. We can therefore conclude that the $q=1$ model is preferred by the EMOND paradigm. 

However, the results of this work seem to be at contention with observations of DF44. A reanalysis of this calculation must be conducted when more UDG velocity dispersions are observed. 

The UDGs serve as a very good test for MOND-like gravity theories and should be studied in more detail. The next step would be to conduct the same analysis for the Virgo cluster and its UDG population.

UDGs are still a relatively new discovery, with limited observations and a small sample size. We predict that more accurate measurements will be made of the velocity dispersions for the UDGs in the near future and with that comes more accurate dynamical mass estimates. It is hard to discuss possible formation scenarios in the context of EMOND as it is still a relatively new theory with limited research conducted. We hope that the take away message of this work is that a possible solution to the mass discrepancy in galaxy clusters in a MOND-like paradigm, EMOND, may also hold the answer to the nature of these UDGs. When two problems have one solution, it warrants further investigation and we hope that EMOND will be investigated further as a result of this.

\section*{ACKNOWLEDGEMENTS}

We would like to thank Anne-Marie Weijmans and Benoit Famaey for general comments on the draft. We would also like to thank Dennis Zaritsky for discussions on the fundamental manifold. AOH is supported by Science and Technologies Funding Council (STFC) studentship (Grant code: 1-APAA-STFC12).

\bibliographystyle{aa} 
\bibliography{UDGbib} 

\appendix
\section{Surface Brightness Conversion}\label{Appendix1}
{  In \cite{UDGDF}, the central surface brightness is given. The FM required the mean surface brightness at the effective radius.} Converting the surface brightness from the value at the centre of the UDG to the mean surface brightness at the effective radius is a simple and standard calculation, that review here for completeness. For a full, more detailed look at the calculation, we refer the reader to \cite{sersic}, where most of the equations below come from. Light profiles are commonly modelled with a S$\acute{ \rm e}$rsic profile. In terms of the surface brightness, $I$, the S$\acute{ \rm e}$rsic profile is
\begin{equation}\label{surfacebrightness}
I(r) = I_{e} + \frac{2.5 b_{n}}{\ln 10} \left[ \left(\frac{r}{r_{e}}\right)^{1/n} - 1 \right]
\end{equation}
\noindent where $n$ parametrises the  S$\acute{ \rm e}$rsic index which describes the shape of the profile and $b_{n}$ is a constant which is defined for each $n$.
\noindent As \cite{UDGDF} quotes the central surface brightness and the FM requires the mean surface brightness at the effective radius, the first step is to solve Eqn \ref{surfacebrightness} for $I_{e}$. All the UDGs in the sample have been modelled with a S$\acute{ \rm e}$rsic index $n=1$. The corresponding $b_{1}$ value is  $\approx 1.678$. Therefore, 
\begin{equation}
I_{e} \approx I_{0} + 1.821.
\end{equation}
\noindent Next we have to transform this value into the average value at the effective radius. The average intensity is defined to be,
\begin{equation}\label{meansurfacebrightness}
\langle {\rm Intensity} \rangle |_{r=r_{e}}\equiv \int^{r_{e}}_{0}\frac{{\rm Intensity(r)}~ 2~\pi~ r~ dr}{\pi r_{e}^{2}}.
\end{equation}
\noindent where the intensity can be transformed into surface brightness via ${I} = 2.5 \log_{10} {\rm (Intensity)}$. Solving Eqn \ref{meansurfacebrightness}, and moving from intensity to surface brightness, we get,
\begin{equation}
\langle I_{e} \rangle = I_{e} - 2.5 \log_{10} \left[\frac{n \exp(b_{n})}{b_{n}^{2n}}\Gamma(2n)\right].
\end{equation}
\noindent Inserting the numbers we arrive at,
\begin{equation}
\langle I_{e} \rangle = I_{0} + 1.821 -0.699
\end{equation}
\noindent where we have expressed the value in terms of the central value of surface brightness. Currently, the mean surface brightness is in units of mags/arcsec$^{2}$ which we need to convert to  L$_{\odot}$/pc$^{2}$. This is done via,
\begin{equation}
I({\rm L_{\odot}/pc^{2}}) = \exp \left[-\frac{\left(I({\rm mags_{\odot}/arcsec^{2}}\right) - M_{\odot} - 21.572 )}{2.5}\right].
\end{equation}
\noindent where $M_{\odot}$ is the solar magnitude in the given band. Therefore,
\begin{equation}
\langle I_{e}\rangle({\rm L_{\odot}/pc^{2}}) = \exp \left[-\frac{I_{0} + 1.821 -0.699 M_{\odot} - 21.572 )}{2.5}\right]
\end{equation}
\noindent where $I_{0}$ is in mags/arcsec$^{2}$. This is the derivation of Eqn \ref{SBprofile1}.
\end{document}